\documentclass[a4paper,11pt]{article}
\pdfoutput=1 
\usepackage{jcappub} 
\usepackage[T1]{fontenc} 
\usepackage{color}
\DeclareMathAlphabet{\mathpzc}{OT1}{pzc}{m}{it}
 
\def\beq{\begin{equation}}
\def\eeq{\end{equation}}
\def\bea{\begin{eqnarray}}
\def\eea{\end{eqnarray}}
\def\kms{~{\rm km\cdot s^{-1}}}
\def\mpc{{\rm Mpc^{-1}}}
\def\ie{{\rm i.e.}}

\def\bn{{\bf n}}

\def\nn{\nonumber\\}
\def\dsum{\displaystyle\sum}
\title{\boldmath Probing beyond-Horndeski gravity on ultra-large scales}
\author[a,b,c]{Didam G.A. Duniya,}\emailAdd{adamsgwazah@gmail.com}
\author[a]{Teboho Moloi,}\emailAdd{tebzanaaka@gmail.com}
\author[d,c,a]{Chris Clarkson,}\emailAdd{chris.clarkson@qmul.ac.uk}
\author[a]{Julien Larena,}\emailAdd{julien.larena@gmail.com}
\author[c,e]{Roy Maartens,}\emailAdd{roy.maartens@gmail.com}
\author[a]{Bishop Mongwane}\emailAdd{astrobish@gmail.com}
\author[a]{and Amanda Weltman}\emailAdd{awelti@gmail.com}

\affiliation[a]{Department of Mathematics \& Applied Mathematics, University of Cape Town, Cape Town 7701, South Africa}%
\affiliation[b]{African Institute for Mathematical Sciences (AIMS), Cape Town 7945, South Africa}%
\affiliation[c]{Department of Physics \& Astronomy, University of the Western Cape, Cape Town 7535, South Africa}%
\affiliation[d]{School of Physics \& Astronomy, Queen Mary University of London, London E1 4NS, UK}%
\affiliation[e]{Institute of Cosmology \& Gravitation, University of Portsmouth, Portsmouth PO1 3FX, United Kingdom}%
\abstract{The beyond-Horndeski gravity has recently been reformulated in the dark energy paradigm---which has been dubbed, Unified Dark Energy (UDE). The evolution equations for the given UDE appear to correspond to a non-conservative dark energy scenario, in which the total energy-momentum tensor is not conserved. We investigate both the background cosmology and, the large-scale imprint of the UDE by probing the angular power spectrum of galaxy number counts, on ultra-large scales; taking care to include the full relativistic corrections in the observed overdensity. The background evolution shows that only an effective mass smaller than the Planck mass is needed in the early universe in order for predictions in the given theory to match current observational constraints. We found that the effective mass-evolution-rate parameter, which drives the evolution of the UDE, acts to enhance the observed power spectrum and, hence, relativistic effects (on ultra-large scales) by enlarging the UDE sound horizon. Conversely, both the (beyond) Horndeski parameter and the kineticity act to diminish the observed power spectrum, by decreasing the UDE sound horizon. Our results show that, in a universe with UDE, a multi-tracer analysis will be needed to detect the relativistic effects in the large-scale structure. In the light of a multi-tracer analysis, the various relativistic effects hold the potential to distinguish different gravity models. Moreover, while the Doppler effect will remain significant at all epochs and, thus can not be ignored, the integrated Sachs-Wolfe, the time-delay and the potential (difference) effects, respectively, will only become significant at epochs near $z\,{=}\,3$ and beyond, and may be neglected at late epochs. In the same vein, the Doppler effect alone can serve as an effective cosmological probe for the large-scale structure or gravity models, in the angular power spectrum---at all $z$.}

\begin{document}
\maketitle
\flushbottom


\section{Introduction}\label{sec:intro}
Observations have shown \cite{Riess:1998,Perlmutter:1999} that the expansion of the universe at recent times is not slowing down, as would be expected from the self-gravity of the galaxies, but is actually accelerating. This mysterious behaviour has been eluding all possible theoretical explanations and available technology. A key problem in cosmology therefore, is to identify the cause of this late-time cosmic accelerated expansion. One explanation, within Einstein's General Relativity theory---which is widely taken as the standard theory of gravity---is that, the acceleration is driven by a non-luminous ``anti-gravity'' agent called Dark Energy (DE) \cite{Copeland:2006wr, Tsujikawa:2010sc, Duniya:2013eta, Duniya:2015nva, Duniya:2015dpa, Duniya:2015ths, Duniya:2016gcf}. However, so far there does not appear to be a fundamental theory for DE. 

Thus, given the lack of a basic understanding of DE, an alternative is that maybe DE is not real, but that the acceleration is rather driven by a relative weakening of gravity at late times on cosmological scales---i.e. a breakdown of general relativity in the infrared---hence, general relativity needs modification. However, this approach has led to the proliferation of Modified Gravity (MG) models \cite{Copeland:2006wr, Tsujikawa:2010sc, Clifton:2011jh}--\cite{Raveri:2019mxg}. (See, particularly, \cite{Clifton:2011jh, Ishak:2018his} for extensive reviews on MG models.) Among the numerous MG models, the \emph{beyond-Horndeski} subclass \cite{Gleyzes:2014rba, Gleyzes:2014dya, Gleyzes:2014qga, Lombriser:2015cla,  Langlois:2015cwa, Akita:2015mho, Crisostomi:2016tcp, Achour:2016rkg, Bellini:2014fua, Sakstein:2016ggl, Brihaye:2016lin, Babichev:2016jom} have received a wide interest of recent, e.g. a so-called ``unified dark energy'' (UDE) model~\cite{Gleyzes:2014rba} has recently been constructed from the beyond-Horndeski gravity.

The given UDE model seeks to combine---in a single description---a broad spectrum of the (well-known) existing models, such as the quintessence models, the scalar-tensor theories and their Horndeski extensions; the $F(R)$ and the Horava-Lifshitz theories, respectively. This approach allows for a unified analysis of cosmological perturbations about a Friedmann-Lemaitre-Robertson-Walker universe, at linear order. Most importantly however, the description of the UDE provides a means for a generalized approach to confront theoretical ideas with observations; the cosmological parameters of the given UDE may be probed and the implication for various models is thus inferred, rather than probing the individual models.

Moreover, the upcoming surveys of the large scale structure---which will span very large cosmic scales, to near the Hubble horizon; reaching high redshifts $z$---posses the potential to provide new information on the nature of DE and MG (and in principle, will be able to test general relativity itself on ultra-large scales). However, in order for these surveys to yield their potential, we need to correct for relativistic effects \cite{Duniya:2015ths, Duniya:2013eta, Duniya:2015nva, Duniya:2015dpa, Duniya:2016gcf, Renk:2016olm, Bonvin:2011bg}--\cite{Andrianomena:2018aad} which naturally surface in the observed overdensity in redshift space. Until recently, the relativistic effects have been disregarded; nonetheless, they are known to become significant on the same scales and redshifts that will be within the reach of the upcoming surveys. Thus, including the relativistic corrections, and understanding their imprint, will be important in the large scale analysis.

In this paper, we investigate galaxy clustering in the UDE model, by probing the angular power spectrum of the observed galaxy source-count overdensity, on ultra-large scales (i.e. larger than the equality scale)---while fully including relativistic corrections. The main goal is to probe the ultra-large-scale imprint of the UDE, in the presence of relativistic effects, on the clustering of galaxies on very large scales; whether the relativistic effects may be important in discriminating DE and MG models. We start by describing the UDE model in Sec.~\ref{sec:UDEM}. In Sec.~\ref{sec:ONCO} we outline the (observed) relativistic galaxy source-count overdensity, and give the observed galaxy angular power spectrum at various redshifts in Sec.~\ref{sec:NumCls}. We probe the UDE in Sec.~\ref{sec:PUDE}: both the background features (subsection~\ref{subsec:UDEbkgd}) and the imprint of the relativistic corrections (subsection~\ref{subsec:IRE}). We conclude in Sec.~\ref{sec:Concl}.


\section{The Unified Dark Energy Model}\label{sec:UDEM}
Here we outline the UDE model, as proposed by \cite{Gleyzes:2014rba}. We rewrite the equations in conformal coordinates---assuming a late-time universe dominated by matter and UDE only.

\subsection{The background equations}
The background energy density $\bar{\rho}_A$ and background pressure $\bar{p}_A$ of UDE ($A \,{=}\, x$) and of matter (henceforth: dark plus baryonic, $A \,{=}\, m$) are related by 
\beq\label{bkg}
\bar{\rho}_x \equiv 3a^{-2}M^2 {\cal H}^2 - \bar{\rho}_m,\quad\quad \bar{p}_x \equiv -a^{-2}M^2(2{\cal H}' + {\cal H}^2) - \bar{p}_m,
\eeq
where $M$ is an effective mass, ${\cal H} \,{=}\, a'/a$ is the comoving Hubble parameter, and $a \,{=}\, a(\eta)$ is the cosmic scale factor; with a prime denoting derivative with respect to conformal time $\eta$. The matter and the UDE background evolution equations, are given by
\beq\label{rhoDots}
\bar{\rho}'_m + 3{\cal H}(1 + w_m)\bar{\rho}_m = 0,\quad\quad \bar{\rho}'_x + 3{\cal H}\left(1 + w_{x,\rm eff}\right)\bar{\rho}_x = 0,
\eeq
where $w_A \,{=}\, \bar{p}_A/\bar{\rho}_A$ is the equation of state parameter of $A$, with 
\beq\label{wx_eff}
w_{x,\rm eff} = w_x - \dfrac{\alpha_M}{3\Omega_x},
\eeq
which denotes an {\it effective} UDE equation of state parameter; $\alpha_M \equiv 2M'/({\cal H} M)$ is the {\em mass evolution rate} parameter---governing the rate of evolution of $M$---and $\Omega_A \,{=}\, \bar{\rho}_A/\bar{\rho}$ is the energy density parameter, with $\bar{\rho}$ being the total background energy density of all the species. 

The evolution of the UDE equation of state parameter is given by
\beq\label{dwxdt}
w'_x = -3{\cal H}\left(1+w_{x,\rm eff}\right)\left(c^2_{ax} - w_x\right),
\eeq
where $c^2_{ax} \equiv \bar{p}'_x / \bar{\rho}'_x$ is the squared adiabatic sound speed of UDE.

\subsection{The perturbation equations}\label{sec:PEqs}
In this subsection we outline the relevant perturbation equations, specifying the constraint and evolution equations. The spacetime metric, is given in Newtonian gauge by
\beq\label{metric}
ds^2 = a(\eta)^2 \left[-(1+2\Phi)d{\eta}^2 + (1-2\Psi) d\vec{x}^2\right],
\eeq
where $\Phi$ and $\Psi$ are the gauge-invariant temporal and spatial metric potentials, respectively, as given by \cite{Gleyzes:2014rba}  (see Appendix \ref{sec:CEqs}). The effective Poisson equation is given by
\beq\label{Pois}
\nabla^2 \Psi = \dfrac{a^2}{2M^2}  \Big\lbrace \dsum_A{\bar{\rho}_A\Delta_A} - \dfrac{\alpha_M}{\Omega_x} \bar{\rho}_x {\cal H}V_x\Big\rbrace, \quad \Delta_A \equiv \delta_A + \dfrac{\bar{\rho}'_A}{\bar{\rho}_A} V_A,
\eeq
where $\Delta_A$ is the effective comoving (energy) density contrast for $A$, and $\delta_A=\delta{\rho}_A/\bar{\rho}_A$ is the Newtonian-gauge energy density contrast, with $V_A$ being the Newtonian-gauge velocity potential. (Note that we take care to use the gauge-invariant comoving overdensity $\Delta_A$, in order to simplify Poisson equation and to define bias properly; moreover this helps avoid any large-scale unphysical artefacts, see e.g. \cite{Bonvin:2011bg, Dent:2008ia}.) The evolution of the spatial metric potential is govern by the total momentum density, given by
\beq\label{PsiDot}
\Psi' + {\cal H}\Phi = -\dfrac{a^2}{2M^2} \dsum_A{q_A},\quad q_A \equiv \left(\bar{\rho}_A+\bar{p}_A\right)V_A,
\eeq
where $q_A \,{=}\, a^{-1}q^{(\rm phys)}_A$ is the effective comoving momentum density; the superscript ``phys'' (henceforth) denotes the physical component---as given by \cite{Gleyzes:2014rba}---i.e. defined with respect to physical time $t$, where $dt \,{=}\, ad\eta$. The metric potentials are related via the constraint equation:
\beq\label{PsiPhi}
\Psi - \Phi = \dfrac{a^2}{M^2} \dsum_A{\sigma_A},
\eeq
where $\sigma_A \,{=}\, a^{-2}\sigma^{(\rm phys)}_A$ is the effective comoving anisotropic stress potential for species $A$.
  
The matter comoving velocity potential and comoving overdensity, respectively, evolve in time according to the equations given by 
\bea\label{VmEvoln}
V'_m + {\cal H}V_m &=&  -\Phi - \dfrac{c^2_{sm}}{1+w_m}\Delta_m - \dfrac{2\nabla^2\sigma_m}{3(1+w_m)\bar{\rho}_m} ,\\ \label{DmEvoln}
\Delta'_m - 3 w_{m} {\cal H}\Delta_m &=& \dfrac{9}{2} {\cal H}^2 (1+w_m)\dsum_A{\Omega_A(1+w_A)\left[V_m - V_A\right]}\nn
&& - \left(1+w_m\right)\nabla^2V_m + \dfrac{2{\cal H}}{\bar{\rho}_m} \nabla^2\sigma_m,
\eea
where $c_{sm}$ is the matter physical sound speed. The physical sound speed of species $A$, given by $c_{sA}$ (which is defined with respect to the rest frame of $A$), measures the propagation speed of the pressure perturbations in the given rest frame. 

Similarly, the UDE comoving velocity potential evolves by
\bea\label{VxEvoln}
V'_x + {\cal H}V_x &=& -\Phi - \dfrac{c^2_{sx}}{1+w_x}\Delta_x -\dfrac{2\nabla^2\sigma_x}{3(1+w_x)\bar{\rho}_x} \nn
&& -\; \dfrac{\alpha_M {\cal H}}{\Omega_x(1+w_x)}\Big[V_x - \dsum_A{\Omega_A(1+w_A)V_A}\Big],
\eea
where $c_{sx}$ is the UDE physical sound speed, given by \cite{Gleyzes:2014dya, Gleyzes:2014qga, Gleyzes:2014rba, Lombriser:2015cla}
\bea\label{c2_sx}
c^2_{sx} &=& -2\dfrac{(1+\alpha_B)^2}{\alpha_K +6\alpha^2_B} \left[ 1 + \alpha_T - \dfrac{1 + \alpha_H}{1 + \alpha_B} \Big(2 + \alpha_M - \dfrac{{\cal H}'}{{\cal H}^2} \Big) - \dfrac{1}{{\cal H}} \Big(\dfrac{1 + \alpha_H}{1 + \alpha_B} \Big)' \right]\nn
&& \quad\quad -\dfrac{(1+\alpha_H)^2}{\alpha_K +6\alpha^2_B} \dfrac{\bar{\rho}_m +\bar{p}_m}{a^{-2}M^2{\cal H}^2},
\eea
with the parameter $\alpha_B$ being dubbed, kinetic {\em braiding}, which measures the kinetic mixing between gravitational and scalar degrees of freedom (in the beyond-Horndeski Lagrangian); the parameter $\alpha_T$ is the {\em tensor speed alteration}, measuring the difference between the (squared) speed $c^2_T \,{=}\, 1 + \alpha_T > 0$ of gravitational waves (or massless gravitons) and the (squared) speed of light; $\alpha_H$ is the (beyond) {\em Horndeski} parameter, which measures the deviation from the Horndeski gravity, and $\alpha_K$ is the {\em kineticity}, which measures the kinetic energy contribution of the scalar field \cite{Gleyzes:2014dya, Gleyzes:2014qga, Gleyzes:2014rba, Lombriser:2015cla, Sakstein:2016ggl, Bellini:2014fua}; where $\alpha_K+6\alpha^2_B >0$. Given the work by \cite{Sakstein:2016ggl}, we have
\beq\label{constraints}
\alpha_B = \alpha_H \Big(1 - 5\dfrac{\Upsilon_2}{\Upsilon_1}\Big),\quad c^2_T = \dfrac{4\alpha^2_H + (1+\alpha_H)\Upsilon_1}{(1+\alpha_B)\Upsilon_1},
\eeq
where cosmological constraints are placed on the parameter governing deviations from Newton's law, $\Upsilon_1 = -0.11^{+0.93}_{-0.67}$, and the parameter governing light bending, $\Upsilon_2 = -0.22^{+1.22}_{-1.19}$.  Clearly we see that, if $\alpha_H \,{=}\, 0$, then $\alpha_B \,{=}\, 0$ and $c^2_T \,{=}\, 1$ (consequently, $\alpha_T \,{=}\, 0$); thus implying that the Horndeski gravity ($\alpha_H \,{=}\, 0$)---as well as general relativity---restricts gravitational waves to propagate with the speed of light. However, in beyond-Horndeski gravity ($\alpha_H \neq 0$), the gravitational waves can propagate either faster ($\alpha_T >0$) or slower ($\alpha_T <0$) than light. 

Moreover, the UDE comoving overdensity evolves by
\bea\label{DxEvoln}
\Delta'_x - 3 w_x {\cal H}\Delta_x &=& \dfrac{9}{2} {\cal H}^2 (1+w_x)\dsum_A{\Omega_A(1+w_A)\left[V_x - V_A\right]} - (1+w_x)\nabla^2V_x +  \dfrac{2{\cal H}}{\bar{\rho}_x} \nabla^2\sigma_x\nn
&+& \dfrac{\alpha_M {\cal H}}{\Omega_x} \Big\lbrace V'_x -\Delta_x + \Big[\dfrac{\alpha'_M}{\alpha_M} -\dfrac{1}{2} (1 + 9w - 2\alpha_M) {\cal H}\Big] V_x\nn
&&\hspace{1.7cm} + \dsum_A{\Omega_A\Big[\Delta_A -\dfrac{\bar{\rho}'_A}{\bar{\rho}_A}V_A -3{\cal H}(1+w_A)V_A\Big]} \Big\rbrace ,
\eea
where \eqref{bkg}--\eqref{DxEvoln} thus constitute the relevant background and perturbations equations. 

By comparing the given equations in this section with the work by e.g.~\cite{Duniya:2015nva}, we see that the UDE essentially corresponds to an interacting DE scenario in which the total energy-momentum tensor is not conserved (there are no $\alpha_M$ terms in all the matter evolution equations, unlike those for UDE; thus $\alpha_M$ induces a self, non-conservative interaction in UDE).


\section{The Observed Source-count Overdensity}\label{sec:ONCO}
For pure source-count surveys, the observed overdensity \cite{Duniya:2015ths, Bonvin:2011bg}--\cite{Durrer:2016jzq}---including all relativistic corrections---along a spatial direction ${\bf n}$ at redshift $z$, is given by 
\beq\label{Delta_n}
\Delta^{\rm obs}_n({\bf n},z) =\; \Delta^{\rm std}_n({\bf n},z) + \Delta^{\rm rels}_n({\bf n},z),  
\eeq
where the \emph{standard} source-count overdensity is given by
\beq\label{Delta_std}
\Delta^{\rm std}_n \;\equiv\; \Delta_{\rm g} - \dfrac{1}{{\cal H}} \partial_r V_\parallel + \int^{r_S}_0{dr \left(r - r_{_S}\right)\dfrac{r}{r_{_S}} \nabla^2_\perp \left(\Phi + \Psi\right)},
\eeq
with $r_{_S} \,{=}\, r(z_{_S})$ being the background comoving radial distance at the source redshift $z_{_S}$, and $V_\parallel \equiv -{\bf n}\cdot{\bf V} \,{=}\, \partial_rV$ is the line-of-sight component of the peculiar velocity---with $V$ being a gauge-invariant velocity potential; ${\cal H} \,{=}\, {\cal H}(z)$ is the comoving Hubble parameter, and $\nabla^2_\perp \,{=}\, \nabla^2 - \partial^2_r - 2r^{-1}\partial_r$ is the square of the image-plane Laplacian. The first and the second terms in the right hand side of \eqref{Delta_std} are the comoving galaxy overdensity and the well-known (Kaiser) redshift distortion term, respectively; the integral constitutes the weak gravitational lensing effect. We have included the lensing in the standard contribution even though it is of course a relativistic effect. This is because we focus on effects on ultra-large scales where the relativistic effects appear at ${\cal O}({\cal H}/k)$ and higher. The relativistic-correction part is given by
\beq\label{Delta_GR}
\Delta^{\rm rels}_n = \Delta^{\rm Doppler}_n + \Delta^{\rm ISW}_n + \Delta^{\rm timedelay}_n + \Delta^{\rm potential}_n,
\eeq
with the various correction terms expressed as follows,
\beq\label{Doppler}
\Delta^{\rm Doppler}_n {\equiv} \Big(b_e - \dfrac{{\cal H}'}{{\cal H}^2}  - \dfrac{2}{r_{_S} {\cal H}}\Big) V_\parallel  + \dfrac{1}{\cal H} \partial_r \left[ \dfrac{c^2_{sm}}{1+w_m}\Delta_m + \dfrac{2\nabla^2\sigma_m}{3(1+w_m)\bar{\rho}_m} \right],
\eeq
where the terms in the square brackets come from the Euler equation \eqref{VmEvoln}, which arise as a result of the redshift perturbation in the volume distortion (see e.g.~\cite{Bonvin:2011bg, Duniya:2015nva, Duniya:2016ibg, Duniya:2015ths, Bonvin:2014owa}); 
\bea\label{ISW}
\Delta^{\rm ISW}_n &{\equiv}& \Big(b_e - \dfrac{{\cal H}'}{{\cal H}^2}  - \dfrac{2}{r_{_S} {\cal H}}\Big) \int^{r_S}_0{dr \left(\Phi' + \Psi' \right) }, \\ \label{timedelay}
\Delta^{\rm timedelay}_n &{\equiv}& \dfrac{2}{r_{_S}}\int^{r_S}_0{dr \left(\Phi + \Psi\right)} ,\\ \label{potential}
\Delta^{\rm potential}_n &{\equiv}& \left(3 - b_e\right){\cal H}V + \dfrac{1}{{\cal H}}\Psi' - 2\Psi - \Big(b_e -1 - \dfrac{{\cal H}'}{{\cal H}^2}  - \dfrac{2}{\bar{r}_{_S} {\cal H}}\Big)\Phi ,
\eea
where $b_e \,{=}\, b_e(z)$ is the galaxy evolution bias~\cite{Duniya:2015ths, Duniya:2016ibg, Jeong:2011as}. In~\eqref{Doppler} we have the Doppler term, which bears the effect of the motion of the source relative to the observer; \eqref{ISW} gives the integrated Sachs-Wolfe (ISW) term, which measures the effect of the phenomenon of losing and gaining energy by signals in propagating through successive potential ``hills'' and ``wells'', from the source to the observer; \eqref{timedelay} gives the time-delay term, which measures the time delay/lag of signals in overcoming potentials of intervening objects along the line of sight, and \eqref{potential} gives the potentials term---including both peculiar-velocity and gravitational potential parts---which contains the potential-difference effect, between the source and the observer.

The galaxy and the matter effective comoving overdensities are related via the linear {\it galaxy bias} $b$ \cite{Duniya:2015ths, Duniya:2016ibg, Jeong:2011as, Bartolo:2010ec, Baldauf:2011bh}, given by $\Delta_{\rm g} \,{=}\, b\Delta_m$.


\section{The Source-count Angular Power Spectrum}\label{sec:NumCls}%

Here we consider the observed source-count overdensity \eqref{Delta_n}, for galaxy surveys; we expand the source-count overdnesity in spherical multipoles, given by
\beq\label{Delta:ell}
\Delta^{\rm obs}_n(\bn,z) = \dsum_{\ell m} {a_{\ell m}(z) Y_{\ell m}(\bn)}, \quad a_{\ell m}(z) = \int{ d^2\bn\, Y^*_{\ell m}(\bn) \Delta^{\rm obs}_n(\bn,z) },
\eeq
where $Y_{\ell m}(\bn)$ are the spherical harmonics and $a_{\ell m}$ are the multipole expansion coefficients, with the asterisk denoting complex conjugate. The total angular power spectrum observed at a source redshift $z_{_S}$ is computed by
\bea\label{Cl_n1}
C_\ell(z_{_S}) &=& \left\langle{\left.\left| a_{\ell m}(z_{_S}) \right.\right|^2 }\right\rangle,\nn 
&=&  \dfrac{4}{\pi^2} \left(\dfrac{43}{50}\right)^2\int{dk\, k^2 T(k)^2 P_{\Phi_p}(k)\Big|f_\ell(k,z_{_S}) \Big|^2 },
\eea
where $T(k)$ is the linear transfer function, $P_{\Phi_p}(k)$ is the primordial power spectrum, and
\bea\label{f_ell}
f_\ell(k,z_{_S}) &=& b(z_{_S}) \tilde{\Delta}_m(k,z_{_S}) j_\ell(kr_{_S}) - j''_\ell(kr_{_S})\dfrac{1}{{\cal H}} \partial_r \tilde{V}^\parallel_m(k,z_{_S}) + j_\ell(kr_{_S})\dfrac{1}{{\cal H}}\tilde{\Psi}'(k,z_{_S}) \nn
&+& \left(3 - b_e\right){\cal H}\tilde{V}_m(k,z_{_S}) j_\ell(kr_{_S}) - \left(b_e -1 - \dfrac{{\cal H}'}{{\cal H}^2}  - \dfrac{2}{r_{_S} {\cal H}}\right) \tilde{\Phi}(k,z_{_S}) j_\ell(kr_{_S}) \nn
&-& 2\tilde{\Psi}(k,z_{_S}) j_\ell(kr_{_S}) + \dfrac{1}{r_{_S}}\int^{r_S}_0{dr\, j_\ell(kr) \left[2  - \dfrac{\left(r - r_{_S}\right)}{r} \ell\left(1+\ell\right) \right] \left(\tilde{\Phi} +\tilde{\Psi}\right)(k,r) } \nn
&+& \left(b_e - \dfrac{{\cal H}'}{{\cal H}^2} - \dfrac{2}{r_{_S} {\cal H}}\right) \left[ j'_\ell(kr_{_S}) \tilde{V}^\parallel_m(k,z_{_S})  + \int^{r_S}_0{dr\, j_\ell(kr) \Big(\tilde{\Phi}' + \tilde{\Psi}' \Big)(k,r) } \right] ,
\eea
where (henceforth) we assume pressureless matter, i.e. $\bar{p}_m \,{=}\, 0 \,{=}\, \delta{p}_m$; thus we have $w_m \,{=}\, 0 \,{=}\, \sigma_m$ and $c^2_{sm} \,{=}\, 0 \,{=}\, c^2_{am}$. For the spherical Bessel function $j_\ell$, we have $j'_\ell(x) \,{=}\, \partial{j}_\ell(x)/\partial{x}$, with $x \,{=}\, kr$. A tilde denotes division by the gravitational potential at the epoch of photon-matter decoupling $z=z_d$, \ie~$\tilde{X}(k,z) \equiv X(k,z)/\Phi_d(k)$ for a given parameter $X$, and~\cite{Duniya:2013eta, Duniya:2015nva, Duniya:2015ths, Dodelson:2003bk}
\beq\label{Phi_d}
\Phi(k,z_d) \;=\; \dfrac{43}{50} \Phi_p(k) T(k) \;\equiv\; \Phi_d(k),
\eeq
where $\Phi_p$ is the primordial gravitational potential; $\tilde{X}$ essentially measures the growth function of the associated parameter. (See e.g.~\cite{Duniya:2013eta, Duniya:2015nva, Duniya:2015ths} for the linear growth functions of $\Delta_m$, $V_m$ and $\Psi$. Note that in the aforementioned references, it is assumed that $\Phi \,{=}\, \Psi$; dropping this assumption leads to the factor $43/50$---in \eqref{Cl_n1} and \eqref{Phi_d}---instead of $9/10$ \cite{Dodelson:2003bk}.) 

Similarly, given \eqref{Delta_std}, the standard angular power spectrum $C^{\rm std}_\ell$ is computed following \eqref{Delta:ell}--\eqref{Phi_d}. In \eqref{f_ell}, we used that on very large scales (which are the scales of interest in this work) $V^\parallel_m \,{=}\, V_\parallel \,{=}\, V^\parallel_{\rm g}$~\cite{Duniya:2015ths, Duniya:2016ibg}, \ie~glaxies flow with the underlying matter---given the homogeneity and isotropy on the very large scales.


\section{Probing the Unified Dark Energy}\label{sec:PUDE}

We initialize all evolutions at the photon-matter decoupling epoch $1+z_d \,{=}\, 10^3 \,{=}\, a(z_d)^{-1}$; using adiabatic initial conditions (see Appendix \ref{sec:AICs}) for the perturbations. We adopt the present-epoch matter density parameter $\Omega_{m0} \,{=}\, 0.3$, and the Hubble constant $H_0 \,{=}\, 67.8\kms\cdot\mpc$ \cite{Ade:2015xua}. Moreover, we use a galaxy bias $b \,{=}\, 1$ and galaxy evolution bias $b_e \,{=}\, 0$ (i.e. where galaxies do not merge with one another); we do this for simplicity as we do not focus on specific surveys.

Although a recent detection of a source in gravitational and electromagnetic radiation showed that the gravitational wave speed $c_{_T} \,{=}\, 1$, it is still possible that $c_{_T}$ varies for higher redshift sources and possibly even with frequency (see e.g.~\cite{Noller:2018wyv}). For generality, and since we are investigating only the qualitative effects of beyond-Horndeski models on galaxy counts, we therefore do not restrict $c_{_T}$ to be unity. We study the behaviour of the background parameters (subsection~\ref{subsec:UDEbkgd}) and the associated large-scale effects of the relativistic corrections in the galaxy source-count angular power spectrum (subsection~\ref{subsec:IRE}).

\subsection{Exploring the UDE background}\label{subsec:UDEbkgd}
For convenience we set $w_{x,\rm eff} \,{=}\, {-1}$, and given \eqref{dwxdt}, it implies $w_x$ is a {\em constant}. Consequently, given \eqref{wx_eff}, we have $\alpha_M \,{\propto}\, \Omega_x$. The advantage of the given choice of $w_{x,\rm eff}$ is that, it allows the recovery of the standard concordance model ($\Lambda$CDM) in the background, at some regimes. Thus for all numerical computations, we use
\beq\label{wxeff_aM}
w_{x,\rm eff} = -1,\quad\quad \alpha_M = \alpha_0 \Omega_x,\quad\quad \alpha_0 < 0.6, 
\eeq
where $\alpha_0$ is a constant and, by using the constraint $w_{x0} \,{<}\, {-0.8}$ at today ($z\,{=}\,0$), it leads to $\alpha_0 \,{<}\, 0.6$. We fix the background at today by choosing the values of $\alpha_0$ so that the UDE evolves to give the same values of $\Omega_{m0}$ and $H_0$. Within constraints~\cite{Sakstein:2016ggl} we adopt $\Upsilon_2 \,{=}\, {-0.131}$ and, given that the density of astrophysical objects decreases radially outwards from the centre so that $\Upsilon_1 \,{>}\, 0$ ($< 0$) implies weakening (strengthening) of gravity~\cite{Sakstein:2016ggl}, we choose $\Upsilon_1 \,{=}\, 0.78$ (to correspond to weakening of gravity on cosmological scales, in the late-time universe). 

\begin{figure}\centering
\includegraphics[scale=0.5]{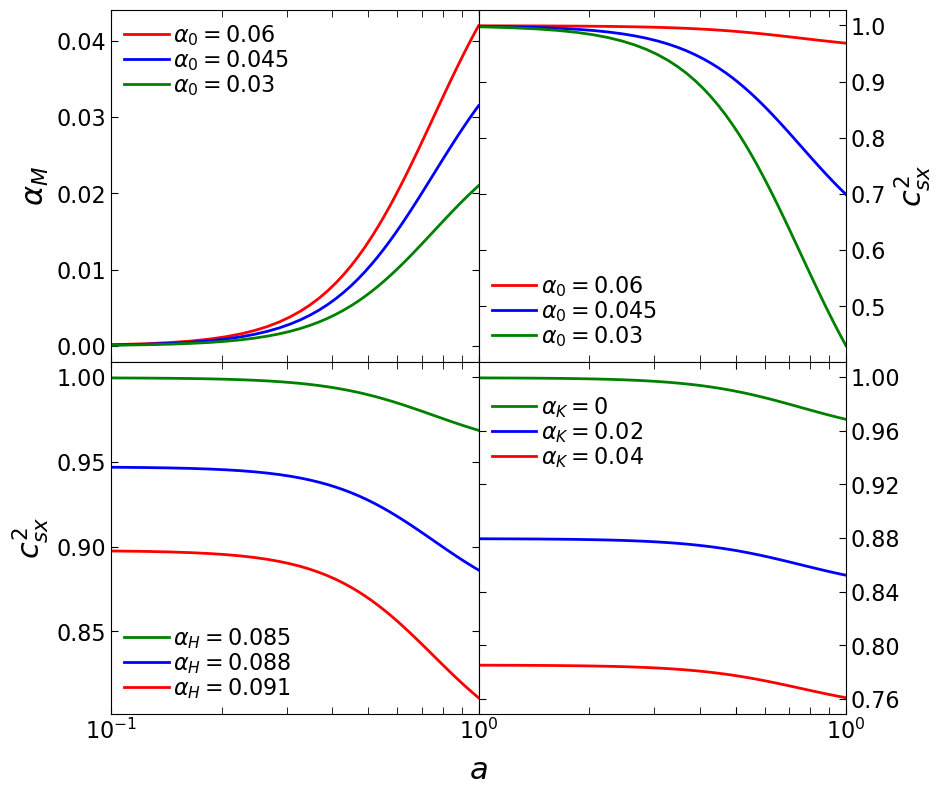}
\caption{{\em Top left:} The plots of the mass-evolution-rate parameter $\alpha_M$, given by \eqref{wxeff_aM}, for the values of the proportionality constant $\alpha_0 \,{=}\, 0.06,\,0.045,\,0.03$---which lead to the values of the UDE equation of state parameter $w_x \,{=}\, {-0.98},\,{-0.985},\,{-0.99}$, respectively. (Note that the evolution of $\alpha_M$ is independent of $\alpha_H$ and $\alpha_K$, which only affect the perturbations.) {\em Top right:} The plots of the corresponding (squared) sound speed $c^2_{sx}$ for the given $\alpha_M$ in the top left panel, with $\alpha_K \,{=}\, 0$ and $\alpha_H \,{=}\, 0.085$. {\em Bottom:} The plots of $c^2_{sx}$ for $\alpha_H \,{=}\, 0.085,\,0.088,\,0.091$ with $\alpha_0 \,{=}\, 0.06$ and $\alpha_K \,{=}\, 0$ ({\em left}), and for $\alpha_K \,{=}\, 0,\,0.02,\,0.04$, with $\alpha_0 \,{=}\, 0.06$ and $\alpha_H \,{=}\, 0.085$ ({\em right}).}\label{fig:alphas} 
\end{figure}

In Fig.~\ref{fig:alphas}, we show the relevant UDE background parameters. We show (top left panel) the behaviour of the effective mass-evolution-rate parameter $\alpha_M$, as a function of the scale factor, a, with $\alpha_0 \,{=}\, 0.03,\,0.045,\,0.06$. (We note that $\alpha_M$ essentially serves to drive the UDE evolution.) We have that at early epochs $z \gtrsim 6$, the beyond-Horndeski theory reduces to general relativity ($\alpha_M \,{\simeq}\, 0$) and the two are identical in the background. However, the beyond-Horndeski gravity deviates ($\alpha_M \,{>}\, 0$) at later epochs $z \,{<}\, 6$. Given the choices in \eqref{wxeff_aM}, the beyond-Horndeski background at $z \,{\gtrsim}\, 6$ resembles $\Lambda$CDM, where the UDE equation of state parameter $w_x \,{\simeq}\, {-1}$; at $z \,{<}\, 6$, the beyond-Horndeski background gradually departs from $\Lambda$CDM (and from a generic general relativistic background), with the UDE having $w_x \,{\neq}\, {-1}$. Thus, the behaviour of $\alpha_M$ may be used to alleviate the well-known coincidence problem---in the sense that, at early epochs $\alpha_M$ vanishes, causing the UDE equation of state parameter to be frozen on the value $w_x \,{=}\, {-1}$ and the UDE mimics the cosmological constant $\Lambda$ (or vacuum energy); at later epochs, $\alpha_M$ grows with time, resulting to $w_x \,{>}\, {-1}$ and the UDE has a positive evolution with an increasing density. (Eventually, the UDE will act to drive the cosmic expansion into an acceleration, at late times $z \,{\lesssim}\, 0.5$.)

The top right panel, Fig.~\ref{fig:alphas}, shows the evolution of the UDE (squared) physical sound speed $c^2_{sx}$, for the given mass parameter $\alpha_M$ (top left panel): for Horndeski parameter $\alpha_H \,{=}\, 0.085$ and kineticity $\alpha_K \,{=}\, 0$. Henceforth, given the values of $\alpha_0$, we adopt values of $\alpha_H$ and $\alpha_K$ such that $c^2_{sx} \,{\leq}\, 1$. Obviously, the behaviour of $c^2_{sx}$ is affected by the evolution of $\alpha_M$, while the late-time amplitude is determined by the values of $\alpha_0$, $\alpha_H$ and $\alpha_K$. We see that when $\alpha_M \,{\simeq}\, 0$ we have $c^2_{sx} \,{\simeq}\, 1$, which is the value predicted by the standard cosmologies. Thus during the given regime, the matter perturbations have similar behaviour as those in $\Lambda$CDM; hence resulting in similar cosmologies---since for any model, the equation of state parameter and the sound speed of the given DE effectively prescribe the cosmology. (By correctly determining these parameters, the associated cosmology is essentially determined.) Similarly, when $\alpha_M \,{>}\, 0$ we have $c^2_{sx} \,{\lesssim}\, 1$ for the given values of $\alpha_0$, $\alpha_H$ and $\alpha_K$. According to the standard cosmologies, only perturbation modes with wavelengths greater than the sound horizon are able to cluster or grow. Thus the larger the sound speed, the more difficult it becomes for the perturbations to cluster. It implies that the growth of $\alpha_M$ at late times should enable the UDE perturbations to cluster---as it decreases the sound horizon.

In the bottom left panel of Fig.~\ref{fig:alphas}, we show the plots of the UDE sound speed for three values of the Horndeski parameter $\alpha_H \,{=}\, 0.085,\, 0.088$ and $0.091$, with $\alpha_0 \,{=}\, 0.06$ and $\alpha_K \,{=}\, 0$. We see that the various values of $\alpha_H$ result in (mostly) constant separations in the amplitude of the UDE sound speed. As the magnitude of the Horndeski parameter increases, the sound speed decreases accordingly, $c^2_{sx} \,{\ll}\, 1$. This implies that, in its strong regime, the beyond-Horndeski gravity will diminish the sound horizon such that the UDE perturbations cluster at low $z$ on sub-Hubble scales. Moreover, in the bottom right panel of Fig.~\ref{fig:alphas} we show the plots of the sound speed for kineticity $\alpha_K \,{=}\, 0,\, 0.02,\, 0.04$, with $\alpha_0 \,{=}\, 0.06$ and $\alpha_H \,{=}\, 0.085$. We see that the behaviour of $c^2_{sx}$ for the given values of $\alpha_K$ is similar to the scenario for $\alpha_H$ (bottom left panel): the amplitude of $c^2_{sx}$ is increased or decreased according to whether the value of $\alpha_K$ is bigger or smaller.

\begin{figure}\centering
\includegraphics[scale=0.5]{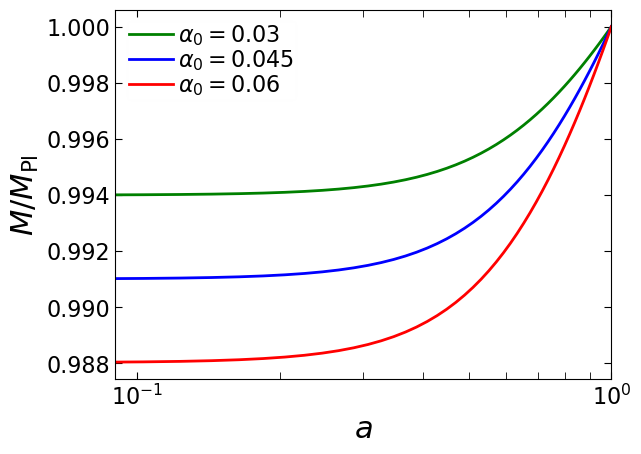}
\caption{The plots of the cosmic evolution of the ratio of the {\em effective} mass $M$ to the standard Planck mass $M_{\rm Pl} \,{=}\, 1/(8{\pi}G)$, with $G$ being the Newton's gravitational constant: for the values $\alpha_0 \,{=}\, 0.03,\,0.045,\,0.06$.}\label{fig:MMp} 
\end{figure}
 
For completeness, in Fig.~\ref{fig:MMp} we show the cosmic evolution of the effective mass $M$---given by the ratio $M/M_{\rm Pl}$---as a function of the scale factor, where $M_{\rm Pl} \,{=}\, (8{\pi}G)^{-1}$ is the standard Planck mass and $G$ is the Newton's gravitational constant (with both the speed of light and the reduced Planck constant as unity). We plot this ratio for various values of $\alpha_0 \,{>}\, 0$. We put a constraint on the effective mass, by setting the present-day value $M(z{=}0) \,{=}\, M_{\rm Pl}$. Thus we have $M \,{<}\, M_{\rm Pl}$ at earlier epochs $z \,{>}\, 0$. We see that as $\alpha_0$ increases, the amplitude of the effective mass decreases at $z \,{>}\, 0$ until matter domination epoch, where it remains constant (for all values of $\alpha_0$). This is understandable given that the dominant cosmic content during this period is pressureless, and unable to induce or drive any growth in mass. It implies that only an effective mass smaller than $M_{\rm Pl}$ is needed in the early universe in order for predictions in the given theory to match current experimental constraints. However, it should be pointed out that, the UDE and/or the cosmic acceleration is not driven by the amplitude of the effective mass, but by its time rate of change---governed by $\alpha_M$. We have that at $z \,{<}\, 1$, the larger the value of $\alpha_0$, the steeper the slope of $M$; consequently, a stronger acceleration. During matter domination, although the amplitude of $M$ is different for different values of $\alpha_0$, the expansion rate remains the same, with $\partial{M}/\partial\eta \,{=}\, 0$ (so that $\alpha_M \,{=}\, 0$).

\subsection{The imprint of relativistic effects}\label{subsec:IRE}
In this section, we probe the observed (relativistic) angular power spectrum. We chose only minimal values of the UDE parameters $\alpha_0$, $\alpha_H$ and $\alpha_K$ for our analysis. We study the large-scale effects of the relativistic corrections \eqref{Doppler}--\eqref{potential} in the angular power spectrum.

\begin{figure}\centering
\includegraphics[scale=0.6]{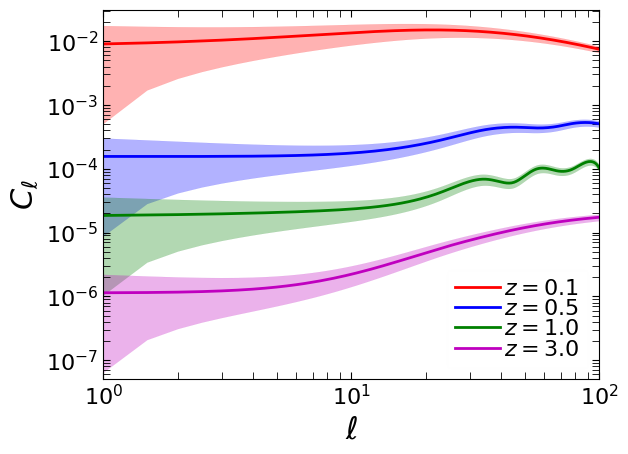} 
\caption{The plots of the full relativistic angular power spectrum $C_\ell$, given by \eqref{Delta_n}--\eqref{Phi_d}, at the source redshifts $z_S \,{=}\, 0.1\, 0.5\, 3$: for fixed values of the parameters $\alpha_0 \,{=}\, 0.06$,  $\alpha_H \,{=}\, 0.085$ and $\alpha_K \,{=}\, 0$. The shaded regions denote cosmic variance at the given redshifts.}\label{fig:Cls_aM}
\end{figure}

In Fig.~\ref{fig:Cls_aM} we show the behaviour of the full relativistic angular power spectrum $C_\ell$ at various redshifts for fixed values of $\alpha_0$, $\alpha_H$ and $\alpha_K$. We plot the $C_\ell$ as a function of multipole $\ell$, at $z_{_S} \,{=}\, 0.1,\, 0.5,\, 1$ and $3$ with, $\alpha_0 \,{=}\, 0.06$, $\alpha_H \,{=}\, 0.085$ and $\alpha_K \,{=}\, 0$. We see that there is a decrease in overall angular power---consequently, in magnitude of galaxy clustering---as redshift increases. These plots agree with already known results for standard cosmologies by previous works (e.g. \cite{Duniya:2013eta, Bonvin:2011bg}, when the scaling factor $\ell(\ell{+}1)/(2\pi)$ on the angular power spectrum is taken into account). We show the extend of the cosmic variance (shaded regions), which is the inherent statistical uncertainty in observing the universe at extreme scales. The cosmic variance for a survey covering a fraction $f_{\rm sky}$ of the sky is given by \cite{Maartens:2012rh, Dodelson:2003bk}
\begin{equation}\label{cosmicVariance}
\sigma_\ell(z) \;=\; \sqrt{ \dfrac{2}{(2\ell+1)f_{\rm sky}} } \; C_\ell(z),
\end{equation}
where $2\ell+1$ counts the number of independent samples used to estimate a given $C_\ell$. As we can see from the plots, it implies that as we move from smaller scales (larger $\ell$) to larger scales (smaller $\ell$)---being the same scales on which relativistic effects are known to be important---cosmological measurements of galaxy clustering will incur bigger uncertainties, at all $z$. Thus, in order for the relativistic effects in the large-scale structure to have significant consequence, they will need to have amplitudes larger than the size of the cosmic variance.

\begin{figure*}\centering
\includegraphics[scale=0.5]{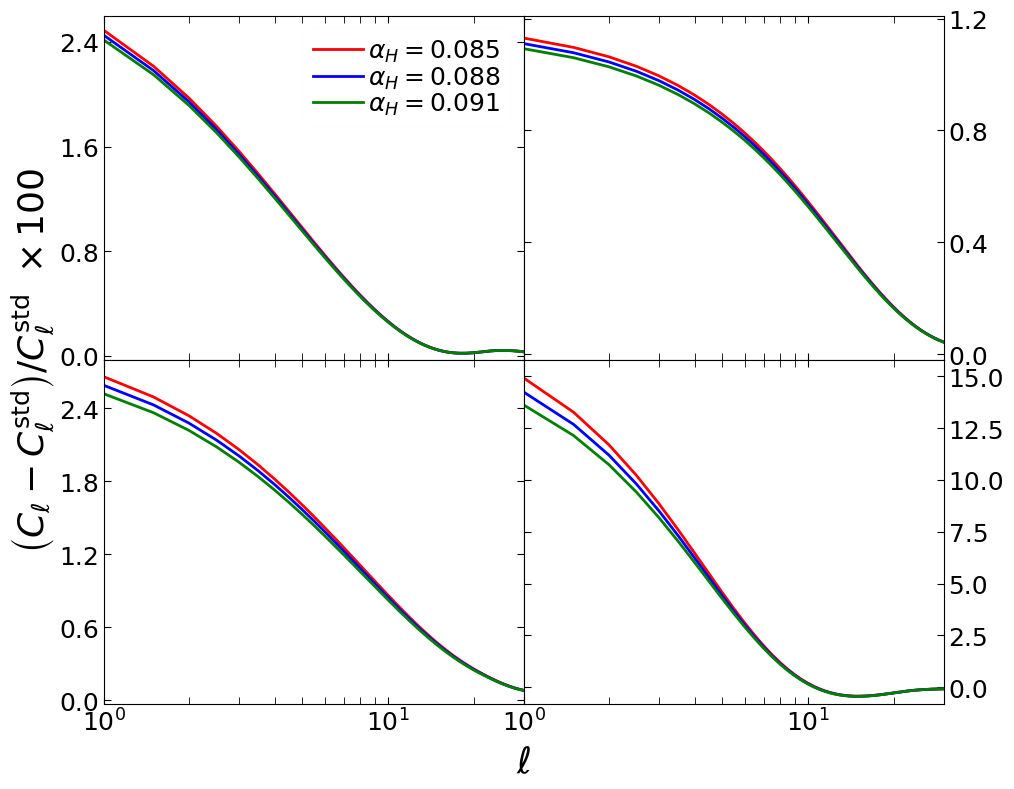}
\caption{The plots of the percentage change in the full relativistic angular power spectrum $C_\ell$ relative to the standard angular power spectrum $C^{\rm std}_\ell$---which is prescribed only by the density, the redshift-distortion and the weak-lensing terms \eqref{Delta_std}---at source redshifts $z_S \,{=}\, 0.1$ (top left), $z_S \,{=}\, 0.5$ (top right), $z_S \,{=}\, 1$ (bottom left) and $z_S \,{=}\, 3$ (bottom right); for $\alpha_H \,{=}\, 0.085,\, 0.088,\,0.091$ with, $\alpha_0 \,{=}\, 0.06$ and $\alpha_K \,{=}\, 0$.}\label{fig:Clsfracs_aH} 
\end{figure*}

In Fig.~\ref{fig:Clsfracs_aH} we show the combined effect of the total relativistic corrections \eqref{Delta_GR} in the angular power spectrum, with respect to different values of the Horndeski parameter $\alpha_H$. We give the plots of the combined relativistic effect as the percentage change in $C_\ell$ relative to the standard angular power spectrum $C^{\rm std}_\ell$---given by~\eqref{Delta_std}---at the epochs $z_{_S} \,{=}\, 0.1,\, 0.5,\, 1$ and $3$; for $\alpha_H \,{=}\, 0.085,\, 0.088,\, 0.091$, with fixed mass-evolution-rate parameter amplitude $\alpha_0 \,{=}\, 0.06$ and vanishing kineticity $\alpha_K \,{=}\, 0$. The results indicate that, at a given redshift, an increase in (the amplitude of) the Horndeski parameter will lead to a suppression of the (combined) relativistic effects in the angular power spectrum on very large scales. However, for a given value of (the amplitude of) $\alpha_H$, the amplitude of the relativistic effects does grow with increasing redshift, for $z \,{\gtrsim}\, 1$. This behaviour also conforms with the results of previous works in the literature on the standard cosmologies.

Results for the different values of $\alpha_K \,{=}\, 0,\, 0.02,\, 0.04$ (with $\alpha_0 \,{=}\, 0.06$ and $\alpha_H\,{=}\, 0.085$), and for the values of $\alpha_0 \,{=}\, 0.03,\, 0.045,\, 0.06$ (with $\alpha_H \,{=}\, 0.085$ and $\alpha_K \,{=}\, 0$) also behave similar to the plots in Fig.~\ref{fig:Clsfracs_aH} except that for $\alpha_0$, the relativistic effects become boosted with larger values of $\alpha_0$; with those of $\alpha_K$ showing similar suppression like in Fig.~\ref{fig:Clsfracs_aH}. Hence, the respective plots of the two scenarios are not shown in this work. Given the results in Fig.~\ref{fig:alphas}, it is understandable to get the enhancement in relativistic effects with larger amplitude of $\alpha_M$: the UDE physical sound speed grows with higher amplitudes of $\alpha_M$; as the sound speed increases, so does the sound horizon---implying that the UDE perturbations are less able to cluster, giving the matter perturbations room to cluster and, consequently, boosting the amplitude of the angular power spectrum. Similarly from Fig.~\ref{fig:alphas}, as the amplitudes of both $\alpha_H$ and $\alpha_K$ increase, the UDE physical sound speed diminishes, accordingly; eventually resulting in the clustering of the UDE perturbations. The growth in the UDE perturbations will suppress the growth in the matter perturbations; thereby diminishing the angular power spectrum---and hence, the relativistic effects. (See Appendix~\ref{sec:stdCls} for further results and comments.)

\begin{figure*}\centering
\includegraphics[scale=0.5]{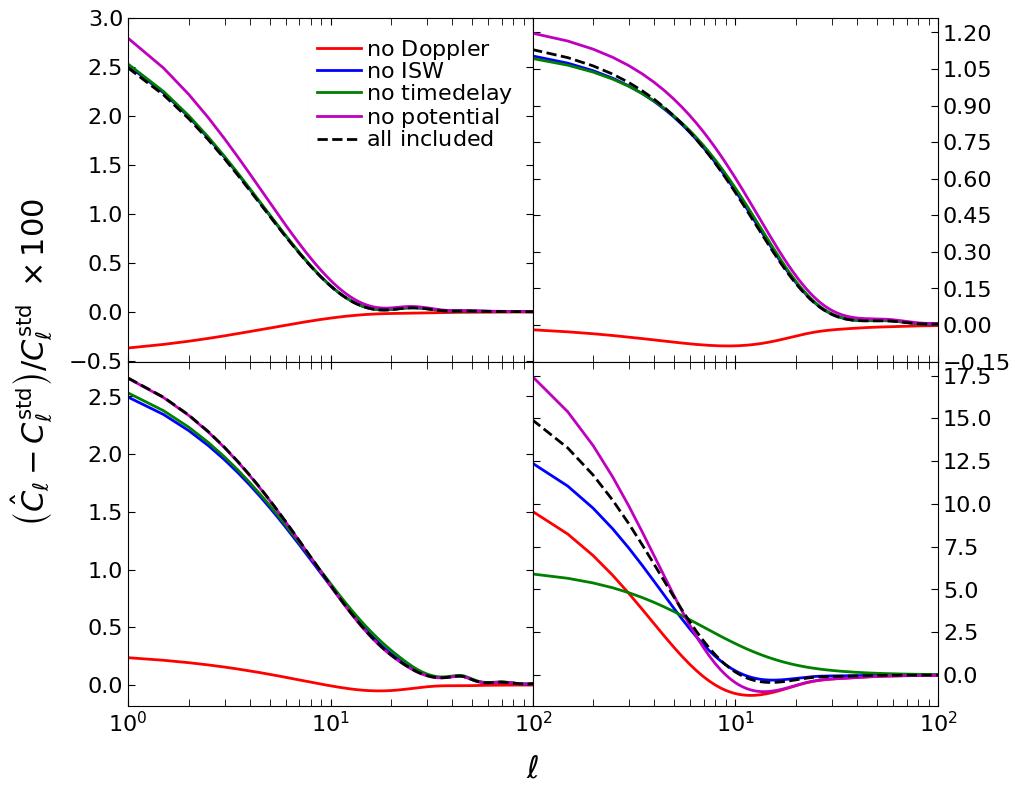}
\caption{The plots of the percentage change in the relativistic angular power spectrum $\hat{C}_\ell$---with the various relativistic terms \eqref{Doppler}--\eqref{potential} individually excluded---relative to the standard agular power spectrum $C^{\rm std}_\ell$. Each line thus, denotes the percentage change when one of the relativistic effect is disregarded, with the black dashed line denoting the change with all effects taken into account (which corresponds to the green lines in Fig.~\ref{fig:Clsfracs_aH}). The panels show the changes at $z_S \,{=}\, 0.1$ (top left), $z_S \,{=}\, 0.5$ (top right), $z_S \,{=}\, 1$ (bottom left) and $z_S \,{=}\, 3$ (bottom right): for the fixed values of $\alpha_0 \,{=}\, 0.06$, $\alpha_H \,{=}\, 0.085$ and $\alpha_K \,{=}\, 0$.}\label{fig:Cls_fracs1_b0_0_03_b1_0_085_b2_0_0}
\end{figure*}

In Fig.~\ref{fig:Cls_fracs1_b0_0_03_b1_0_085_b2_0_0} we examine the individual effect of each relativistic-correction term. We do so by first computing \eqref{Cl_n1}, with all the relativistic terms \eqref{Doppler}--\eqref{potential} included, to obtain $C_\ell$; then repeating the same computation four more times, but each time, we exclude one of the relativistic terms, to obtain $\hat{C}_\ell$. We compute the fractional change in $\hat{C}_\ell$ relative to the standard angular power spectrum $C^{\rm std}_\ell$, accordingly. Thus in Fig.~\ref{fig:Cls_fracs1_b0_0_03_b1_0_085_b2_0_0}, we plot the percentage change in $\hat{C}_\ell$ (which has the various relativistic terms individually ignored, one at a time) relative to $C^{\rm std}_\ell$, for each of the excluded relativistic term, at different redshifts. We show the change in $C_\ell$---which has all the relativistic terms included---relative to $C^{\rm std}_\ell$, as a dashed black line, in comparison to the changes when the various relativistic terms are excluded, as solid coloured lines. We observe that when the potentials (difference) term~\eqref{potential} is excluded (magenta line), the percentage change in the angular power spectrum rises above that which has all the relativistic terms included (dashed black line) and, when the potential term is included, the percentage change in the angular power spectrum falls below that with all the relativistic terms included. That is we have that, including the potentials-effect term decreases the amplitude of the angular power spectrum on very large scales and, vice versa, ignoring this term boosts the angular power spectrum amplitude. Thus this implies that the given relativistic effect has a negative contribution in the angular power spectrum (at all $z$). 

At $z\,{\lesssim}\,1$, the time-delay term~\eqref{timedelay} is relatively insignificant. Neither ignoring the time-delay effect (green line) nor including it (dashed black line) makes any difference; indicating that time delay only makes a negligible contribution in the angular power spectrum at low $z$. At redshifts $z\,{>}\,1$, the contribution of the time-delay effect gradually become substantial---indicated by the separation between the green line and the dashed black line. Moreover, we see that the green line drops down below the dashed black line, indicating that the time-delay effect has a positive contribution in the angular power spectrum on very large scales, at high $z$: since by ignoring time delay, we get a drop in power. This is understandable since the time-delay effect is given by an integral, whose interval---determined by the comoving distance along the line of sight---increases with increasing $z$ (as $z$ decreases, the line-of-sight comoving distance eventually approaches zero). Similarly for the case when the ISW term~\eqref{ISW} is excluded (blue line): the ISW effect only becomes substantial at $z\,{\gtrsim}\,3$ and, appears to be the dominant relativistic effect at these redshifts. At all redshifts, when the Doppler term~\eqref{Doppler} is ignored (red line), the percentage change falls below that which has all the relativistic terms included (dashed black line)---indicating that the Doppler effect has a positive contribution in the angular power spectrum, at all $z$. Moreover, the Doppler effect remains significant at all $z$ and, is the dominant effect at $z\,{<}\,3$ (subdominant at $z\,{\gtrsim}\,3$).

\begin{figure*}\centering
\includegraphics[scale=0.5]{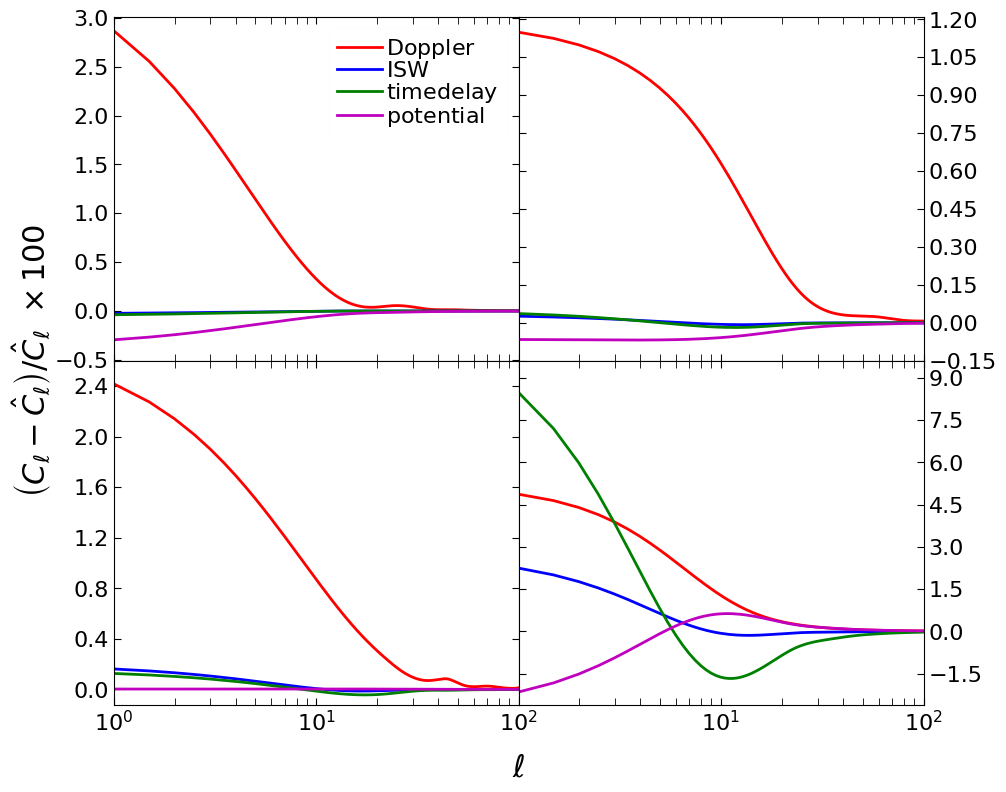}
\caption{The plots of the percentage change in the full relativistic angular power spectrum $C_\ell$ relative to $\hat{C}_\ell$ (as in Fig.~\ref{fig:Cls_fracs1_b0_0_03_b1_0_085_b2_0_0}). Thus each line indicates the actual large-scale effect (in percentage) owing solely to the specified relativistic corrections, in the galaxy source-count angular power spectrum. The panels show the effects at $z_S \,{=}\, 0.1$ (top left), $z_S \,{=}\, 0.5$ (top right), $z_S \,{=}\, 1$ (bottom left) and $z_S \,{=}\, 3$ (bottom right): for the values of $\alpha_0 \,{=}\, 0.06$, $\alpha_H \,{=}\, 0.085$ and $\alpha_K \,{=}\, 0$.}\label{fig:Cls_fracs2_b0_0_03_b1_0_085_b2_0_0} 
\end{figure*}

Essentially, there are two key features to note in Fig.~\ref{fig:Cls_fracs1_b0_0_03_b1_0_085_b2_0_0}. The first is, the position of the solid line with respect to the dashed black line---whether it is above or below. A line rising above the dashed line indicates that the associated relativistic effect has a negative contribution in the angular power spectrum and, conversely, by falling below the dashed black line a solid line indicates that the associated relativistic effect has a positive contribution: since ignoring the relativistic term leads to the boosting of the amplitude of the angular power spectrum, while including the term results in a diminished amplitude. When a solid line coincides with the dashed line, it suggests that the associated relativistic effect is relatively insignificant (since whether or not it is included, the amplitude of the angular power spectrum remains unchanged). The second feature to note is, the separation between a solid line and the dashed black line---the size of which indicates the relative significance of the associated term, with respect to the rest of the relativistic effects. 

In general, the results show that ignoring the potentials term \eqref{potential} will lead to an overestimation of the large-scale power spectrum and/or the combined relativistic effect. Conversely, neglecting the Doppler term \eqref{Doppler} will lead to a significant underestimation of the overall relativistic effects. (It should be pointed out that larger values of $\alpha_0$ can force the ISW and the time-delay effects, respectively---along with the potentials effect---to have negative contribution in the angular power spectrum.) Unlike the Doppler term, both the ISW term \eqref{ISW} and the time-delay term \eqref{timedelay} may be ignored at $z \,{\leq}\,1$ without resulting in any significant deviations. At $z\,{\gtrsim}\,3$, all the effects gradually become significant and can no longer be ignored.

\begin{figure*}
\includegraphics[scale=0.494]{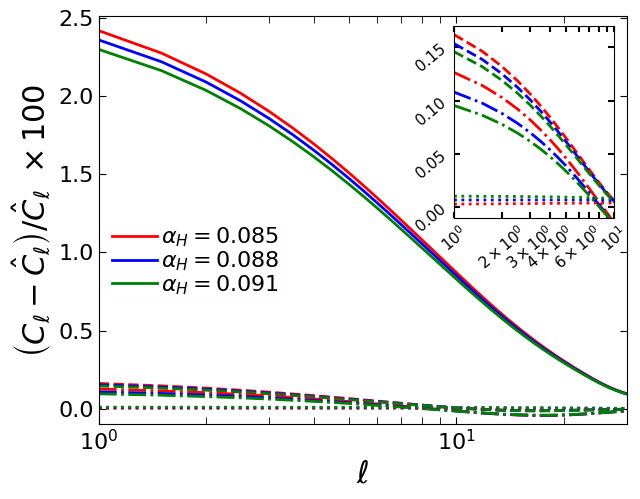}  \includegraphics[scale=0.53]{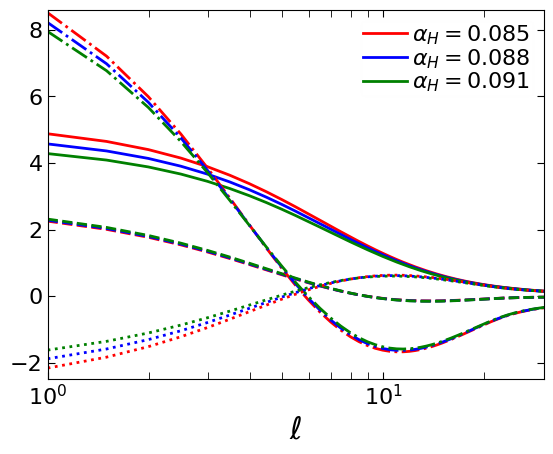}
\caption{The plots of the percentage change owing to the various relativistic effects---for different values of the Horndeski parameter $\alpha_H$: 0.085, 0.088 and 0.091. The \emph{left} panel shows the results at $z_{_S}\,{=}\,1$ and the \emph{right} panel shows the results at $z_{_S}\,{=}\,3$. The relativistic terms are individually dropped (one at a time) in the total angular power spectrum $\hat{C}_\ell$, and the fractional change is calculated accordingly from the total angular power spectrum $C_\ell$ which has all the relativistic terms included. The different line styles denote the particular relativistic effects: Doppler (solid), ISW (dashed), time delay (dot-dashed) and potentials (dotted).}\label{fig:Cls_fracs_a_H} 
\end{figure*}

In Fig.~\ref{fig:Cls_fracs2_b0_0_03_b1_0_085_b2_0_0} we further probe the individual effect of the various relativistic terms~\eqref{Doppler}--\eqref{potential} in the angular power spectrum. Unlike in Fig.~\ref{fig:Cls_fracs1_b0_0_03_b1_0_085_b2_0_0} where we considered the changes relative to the standard angular power spectrum $C^{\rm std}_\ell$, here we compute the various large-scale changes in the total angular power spectrum $C_\ell$ (which has all relativistic terms included) relative to the total angular power spectrum $\hat{C}_\ell$ (which has the various relativistic terms individually excluded). Thus in Fig.~\ref{fig:Cls_fracs2_b0_0_03_b1_0_085_b2_0_0} we give the percentage change in the total relativistic angular power spectrum, relative to itself---after one of the relativistic terms \eqref{Doppler}--\eqref{potential} is excluded. In the figure, we plot the change for each relativistic term being ignored, at the epochs $z_{_S} \,{=}\, 0.1,\, 0.5,\, 1$, and $3$; with $\alpha_0 \,{=}\, 0.06$, $\alpha_H \,{=}\, 0.085$ and $\alpha_K \,{=}\, 0$. The plots therefore give the true individual effect of the various relativistic terms. The key features to note in Fig.~\ref{fig:Cls_fracs2_b0_0_03_b1_0_085_b2_0_0} are, (i) the sign of the percentage change, which indicates the kind of contribution of the associated relativistic term in the angular power spectrum---a positive amplitude corresponds to a positive contribution and, a negative amplitude corresponds to a negative contribution; (ii) the relative separation of the various lines from the zero line: indicating the relative significance of the associated relativistic term. 

At all redshifts, when the potentials term~\eqref{potential} is ignored, the resulting total angular power spectrum $\hat{C}_\ell$ has larger amplitude on very large scales than the total angular power spectrum $C_\ell$ containing all the relativistic terms: the potentials effect (magenta line) diminishes the relativistic angular power spectrum on very large scales. This is shown by the negative values of the change amplitude, which agrees with the results in Fig.~\ref{fig:Cls_fracs1_b0_0_03_b1_0_085_b2_0_0}. We see that indeed the Doppler effect (red line)---although subdominant at $z\,{\gtrsim}\,3$ and, dominant at $z\,{<}\,3$---remains significant at all redshifts. This is understandable since this effect only pertains the relative peculiar velocity between the source and the observer; at all redshifts, sources are in constant peculiar motion owing to surrounding gravitational instabilities. Although the Doppler effect becomes subdominant (only to time-delay effect) at $z\,{\gtrsim}\,3$, its amplitude at these redshifts is larger than that at $z\,{<}\,3$ and, the Doppler effect increases with increasing redshift---as do the other relativistic effects. Both the ISW effect (blue line) and the time-delay effect (green line) remain subdominant and relatively insignificant at $z\,{<}\,1$, like in Fig.~\ref{fig:Cls_fracs1_b0_0_03_b1_0_085_b2_0_0}. In general, we observed that the contribution of all the relativistic effects grow with increasing redshifts at $z\,{\gtrsim}\,1$---as indicated by the growing amplitude of the percentage changes---corroborating already established results in the literature.

\begin{figure*}
\includegraphics[scale=0.494]{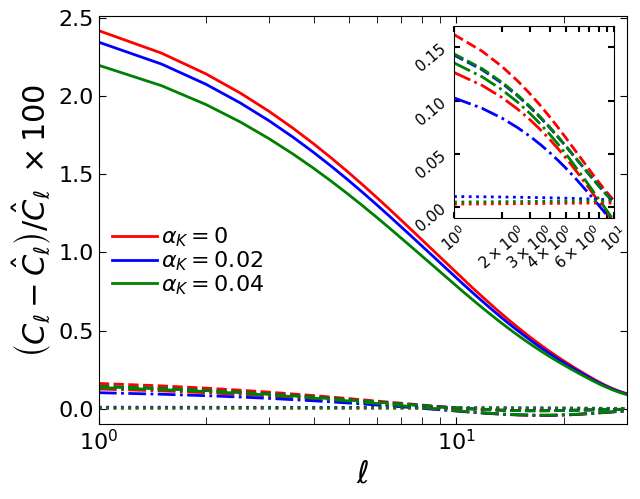}  \includegraphics[scale=0.53]{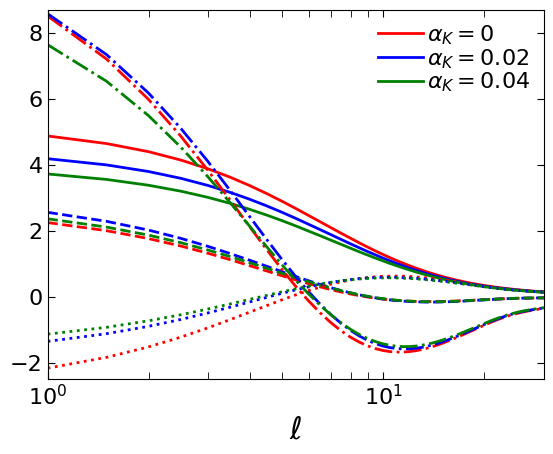}
\caption{The plots of the percentage change owing to the various relativistic effects---for different values of the kineticity $\alpha_K$: 0, 0.02 and 0.04. All notations are as in Fig.~\ref{fig:Cls_fracs_a_H}.}\label{fig:Cls_fracs_a_K} 
\end{figure*}

In Fig.~\ref{fig:Cls_fracs_a_H} we illustrate the large-scale effect of the Horndeski parameter $\alpha_H$ on the relativistic effects, in the angular power spectrum. We repeat the analysis of Fig.~\ref{fig:Cls_fracs2_b0_0_03_b1_0_085_b2_0_0}, except that here we vary $\alpha_H$; fixing $\alpha_K \,{=}\, 0$ and $\alpha_0 \,{=}\, 0.06$. We give the plots of the various relativistic effects in percentage, on large scales---at both $z_{_S}\,{=}\,1$ (left panel) and at $z_{_S} \,{=}\, 3$ (right panel), respectively---for $\alpha_H \,{=}\, 0.085,\, 0.088$ and $0.091$. Overall, the Doppler effect (solid lines) remain highly dominant at $z$ = 1, and gradually become subdominant at $z\,{>}\,1$. We see that the time-delay effect (dot-dashed lines) only becomes significant and dominant at higher redshifts $z\,{\gtrsim}\,3$. The ISW effect (dashed lines) become more significant at $z \,{\gtrsim}\, 3$, yet subdominant---with time delay being the leading effect. Although the large-scale effect of the potentials gradually become significant at $z\,{\gtrsim}\,3$, the associated cosmological contribution remain a negative one. At $z \,{\leq}\, 1$, we have that an increasing $\alpha_H$ will result in the suppression of the Doppler, the ISW and the time-delay effects, respectively; on the other hand, the potentials effect becomes enhanced, but having negative amplitude (at all $z$). At higher redshifts $z\,{\sim}\,3$, although the ISW effect remains positive, it becomes enhanced (as is the potentials effect) with increasing $\alpha_H$; the Doppler and the time-delay effects remain positive and, become suppressed with increasing $\alpha_H$. Moreover, given that the Doppler, the ISW and the time-delay terms, respectively have positive effects, this implies that omitting any of these terms will lead to a decrease in the amplitude of the total angular power spectrum (and the overall relativistic effect), on very large scales. On the other hand, excluding the potentials term will lead to an enhancement in amplitude of the total angular power spectrum on very large scales. 

Similarly, in Fig.~\ref{fig:Cls_fracs_a_K} we illustrate the large-scale effect of the kineticity $\alpha_K$ on the relativistic effects, in the angular power spectrum. We have that an increasing $\alpha_K$ will lead to the suppression of the Doppler effect at all redshifts and, the potentials effect becomes enhanced. On the other hand, increasing or decreasing the amplitude of $\alpha_K$ will cause the ISW and the time-delay effects, respectively, to oscillate: a consistent increase or decrease in $\alpha_K$ leads to an enhancement and then a sudden suppression, or vice versa. This may be owing to our choice of $\alpha_K$ and $\alpha_H$---as absolute constants---which may result in the UDE becoming stiff, being restrained from seeking its natural evolution. Otherwise, the behaviour and explanation of the various relativistic effects here follow from the discussion under Fig.~\ref{fig:Cls_fracs_a_H}.

\begin{figure*}
\includegraphics[scale=0.494]{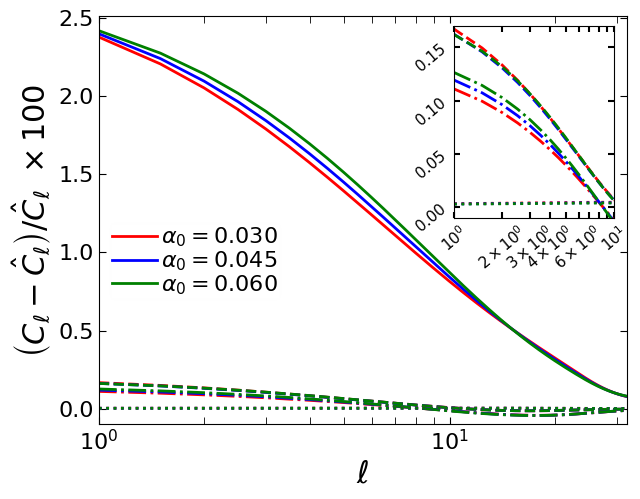}  \includegraphics[scale=0.53]{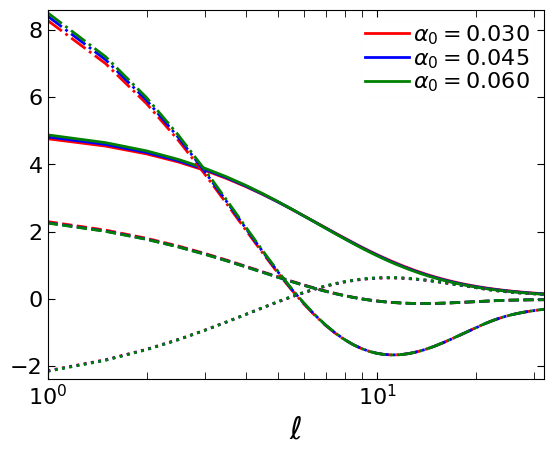}
\caption{The plots of the percentage change owing to the various relativistic effects---for different values of the mass-parameter amplitude $\alpha_0$: 0.03, 0.045 and 0.06. All notations are as in Fig.~\ref{fig:Cls_fracs_a_H}.}\label{fig:Cls_fracs_a_0} 
\end{figure*}

For completeness, in Fig.~\ref{fig:Cls_fracs_a_0} we show the large-scale effect of the mass-parameter amplitude $\alpha_0 \,{=}\, \alpha_M/\Omega_x$ on the relativistic effects, in the angular power spectrum. Although the values of the UDE parameters we chose are only representative, the results suggest that the relativistic effects are relatively less sensitive to any amplitude variations in $\alpha_M$ than in either $\alpha_H$ or $\alpha_K$. The changes for the different values of $\alpha_0$ mostly coincide with one another in the various relativistic effects. Moreover, unlike for $\alpha_K$ where consistent changes in value lead to fluctuations in the ISW and the time-delay effects (see Fig.~\ref{fig:Cls_fracs_a_K}), a consistent change in $\alpha_H$ or $\alpha_0$ leads to a consistent change in all the relativistic effects: growth in $\alpha_H$ gives suppression---except for the potentials effect which is enhanced---and, vice versa (see Fig.~\ref{fig:Cls_fracs_a_H}); growth in $\alpha_0$ tends to enhance the relativistic effects and, vice versa. Both the Doppler and the potentials effects, respectively, respond consistently to the changes in $\alpha_0$, $\alpha_H$ and $\alpha_K$ at all redshifts: the Doppler effect becomes suppressed with increasing $\alpha_H$ and $\alpha_K$ and, becomes enhanced with increasing $\alpha_0$; conversely, the potentials effect becomes enhanced with increasing $\alpha_H$ and $\alpha_K$ and, appears insensitive to the changes in $\alpha_0$ (for the given values). 

\begin{figure}\centering
\includegraphics[scale=0.6]{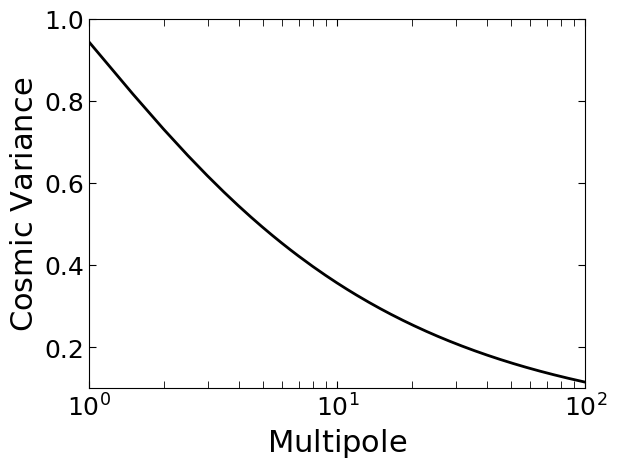} 
\caption{The plot of the cosmic variance, as a ratio with the angular power spectrum---with regards to the fractionational changes in the angular power spectrum.}\label{fig:CV}
\end{figure}

In Fig.~\ref{fig:CV}, we show the cosmic variance associated with the results in Figs.~\ref{fig:Clsfracs_aH}--\ref{fig:Cls_fracs_a_0}. Note that the results in Figs.~\ref{fig:Clsfracs_aH}--\ref{fig:Cls_fracs_a_0} are given as percentages; in order to compare these results to the cosmic variance~\eqref{cosmicVariance} we must also multiply the plots in Fig.~\ref{fig:CV} by $100$. If the percentage change is below the cosmic variance (${\times}\, 100$, as given by Fig.~\ref{fig:CV}), then the associated relativistic effect is not measurable. This is a definite and unambiguous  statement, since we neglect noise in this analysis and so the cosmic-variance error is what is obtainable with a perfect experiment. Thus, as shown by the figures, it implies that in a universe with UDE all the relativistic effects will be subsumed by cosmic variance in the measurements of the large-scale structure. 

It is already well known that in general relativity, the ultra-large scale relativistic effects in the power spectrum are below cosmic variance and therefore undetectable \cite{Alonso:2015uua}. This feature does not depend on the Einstein field equations, and will also persist in metric modified gravity theories. However, if two or more tracers of the matter distribution in the same volume of the Universe are used (the multi-tracer method), this suppresses cosmic variance and the relativistic effects can be detected \cite{Fonseca:2015laa, Alonso:2015sfa, Witzemann:2018cdx}. Once again, this feature will persist in modified gravity. The various relativistic effects may also be isolated (individually) on ultra-large scales via the angular correlation function \cite{Bonvin:2014owa, Bonvin:2015kuc, Tansella:2017rpi, Franco:2018yag, Andrianomena:2018aad}.
  
In the light of a multi-tracer analysis in particular, the various relativistic effects hold the potential to distinguish different gravity models---given the fact that the lines in e.g. Figs.~\ref{fig:Cls_fracs_a_H}--\ref{fig:Cls_fracs_a_0} spread out with the different values of the UDE parameters ($\alpha_H$, $\alpha_K$ and $\alpha_0$) for a particular relativistic effect. The Doppler effect appears to be the most responsive of the relativistic effects, especially to $\alpha_H$ and $\alpha_K$. This suggests that the Doppler effect, in the observed angular power spectrum, can be used alone as an effective cosmological probe for the large scale structure, particularly at $z\,{\leq}\,1$.


\section{Conclusion}\label{sec:Concl}
We have presented a broad analysis of the beyond-Horndeski gravity---as recently reformulated as, Unified Dark Energy (UDE). The evolution equations for the given UDE appear to correspond to a non-conservative dark energy scenario, in which the total energy-momentum tensor is not conserved. We investigated both the background cosmology and, the imprint of relativistic effects in the large-scale structure, by probing the angular power spectrum of galaxy relativistic source counts, on very large scales.

Firstly, we study the behaviour of the UDE background parameters. The effective mass-evolution-rate parameter $\alpha_M$, which drives the evolution of the UDE, can be set to recover the value of the Planck mass at the present epoch. We found that as the amplitude of $\alpha_M$ increases, the amplitude of the UDE effective mass decreases. Moreover, for a given amplitude of $\alpha_M$, the UDE effective mass diminishes with increasing $z\,{>}\,0$ until the matter domination epoch, where it remains constant regardless of the amplitude of $\alpha_M$. We attributed this constancy in amplitude to the fact that matter, being the dominant cosmic component during this period, has zero pressure; consequently, it is unable to induce any growth in mass. Also, the decrease in amplitude with increasing $z$ implies that only an effective mass smaller than the Planck mass is needed in the early universe in order for predictions in the given theory to match current experimental constraints.

Furthermore, by choosing parameters so that the UDE physical sound speed never exceeds unity throughout the cosmic evolution history, our results showed that the behaviour of UDE physical sound speed is strongly governed by the evolution of the mass parameter $\alpha_M$. During the matter epoch, $\alpha_M$ vanishes, owing to the UDE effective mass being constant, resulting in the sound speed becoming unity (and decreases as we move towards the present epoch)---which is the value typical in the standard cosmologies---with the UDE equation of state parameter approaching negative one ($w_x\,{\simeq}\, {-}1$): the value in $\Lambda$CDM. Thus during this regime the matter perturbations will have similar behaviour as those in $\Lambda$CDM. Moreover, we found that the (beyond) Horndeski parameter $\alpha_H$ acts to diminish the amplitude of the sound speed. The larger the amplitude of $\alpha_H$, the smaller the sound speed. Similarly, the kineticity $\alpha_K$ induces the same kind of effect on the sound speed. On the other hand, the mass parameter rather induces growth in the sound speed, with larger amplitude of the mass parameter: the larger the amplitude of $\alpha_M$, the larger the sound speed.

We compute the galaxy source-count angular power spectrum, considering mainly the very large (linear) scales. We took account of the full, known ultra-large scale relativistic corrections---the Doppler, the ISW, the time-delay and the potentials (difference) corrections, respectively---in the observed overdensity. We found that the (combined) relativistic effects become boosted with larger amplitudes of $\alpha_M$. This happens by the fact that $\alpha_M$ enhances the UDE physical sound speed and, as the sound speed increases, so does the sound horizon: implying that the UDE perturbations are less able to cluster, therefore allowing the matter perturbations to grow. Consequently, the amplitude of the angular power spectrum---and hence, the relativistic effects---becomes boosted. Conversely, both $\alpha_H$ and $\alpha_K$ act to diminish the relativistic effects. As the amplitudes of $\alpha_H$ and $\alpha_K$ increase, the UDE physical sound speed diminishes, accordingly. This eventually results in the clustering of the UDE perturbations; consequently, suppressing the growth in the matter perturbations and, hence, diminishing the angular power spectrum.

Our results showed that the potentials effect has a negative contribution in the angular power spectrum at all epochs and, is insignificant at low redshifts $z\,{<}\,1$. The time-delay effect, being an integral effect, is insignificant in the angular power spectrum at low $z\,{<}\,1$. At high $z\,{>}\,1$, the contribution of the time-delay effect gradually become substantial, with a positive contribution in the angular power spectrum on very large scales. Similarly, the ISW effect only becomes substantial at high $z\,{\gtrsim}\,3$. At all redshifts, the Doppler effect has a positive contribution in the angular power spectrum; remains significant at all epochs and, is the dominant effect at $z\,{<}\,3$ (it becomes subdominant at $z\,{\gtrsim}\,3$). Thus neglecting the Doppler effect will lead to a significant underestimation of the relativistic effects in the angular power spectrum. Conversely, both the ISW and the time-delay effects, may be ignored at low $z\,{\leq}\,1$ without resulting in any significant deviations in the relevant cosmological parameters. However, at high $z\,{\gtrsim}\,3$, all the relativistic effects become significant and can no longer be ignored. Excluding any of, the Doppler effect, the ISW effect or the time-delay effect, will lead to a decrease in the amplitude of the observed angular power spectrum, on very large scales; conversely, excluding the potentials effect will lead to an enhancement.

At late epochs $z \,{\leq}\, 1$, the Horndeski parameter acts to suppress the Doppler, the ISW and the time-delay effects. On the other hand, the potentials effect becomes enhanced, at all epochs, but having negative amplitude. At high $z\,{\gtrsim}\,3$, although the ISW effect remains positive, it becomes enhanced---as is the potentials effect---with increasing amplitude of the Horndeski parameter. The Doppler and the time-delay effects remain positive and, become suppressed with increasing amplitude of the Horndeski parameter. We also found that a growth in the amplitude of the kineticity will lead to the suppression of the Doppler effect at all epochs and, will enhance the potentials effect. A change in the amplitude of the kineticity will cause the ISW and the time-delay effects, to oscillate in growth: a consistent increase or decrease in the amplitude of the kineticity will lead to an enhancement and then a sudden suppression, or vice versa. However, both the Doppler and the potentials effects, respond consistently to the changes in all the UDE parameter at all redshifts: the Doppler effect becomes suppressed with growing kineticity and Horndeski parameter and, becomes enhanced with growing mass-evolution-rate parameter. The potentials effect becomes enhanced with growing kineticity and Horndeski parameter.

A multi-tracer analysis will be needed to detect the relativistic effects in the large-scale structure, in a universe govern by beyond-Horndeski gravity. If two or more tracers of the matter distribution in the same volume of the Universe are used (the multi-tracer method), this suppresses cosmic variance and the relativistic effects can be detected. Thus in the light of a multi-tracer analysis, the various relativistic effects hold the potential to distinguish different gravity models. Moreover, the Doppler effect alone can be used as an effective cosmological probe for the large scale structure and/or gravity models at late epochs $z\,{\leq}\,1$, in the observed angular power spectrum.

\acknowledgments{We thank Obinna Umeh for useful comments. This work was carried out with financial support from (i) the government of Canada's International Development Research Centre (IDRC) and, within the framework of the AIMS Research for Africa Project, and (ii) the South African Square Kilometre Array Project and the South African National Research Foundation. CC was supported by STFC Consolidated Grant ST/P000592/1. RM was supported by STFC Consolidated Grant ST/N000668/1, and by the South African Radio Astronomy Observatory (SARAO) and the National Research Foundation (Grant No. 75415). A. Weltman gratefully acknowledges support from the South African Research Chairs Initiative of the Department of Science and Technology and the National Research Foundation of South Africa.}

\appendix

\section{The Standard Angular Power Spectra}\label{sec:stdCls}
\begin{figure}\center
\includegraphics[scale=0.4]{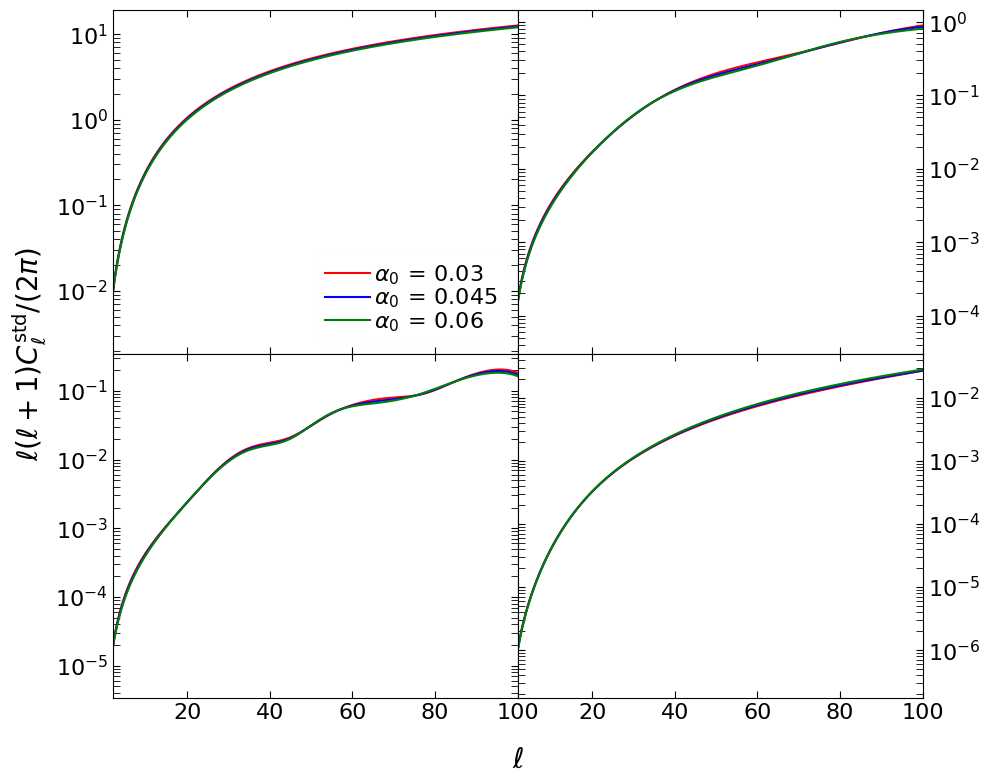}
\caption{The plots of the standard angular power spectrum, as obtained from \eqref{Delta_std}, for the values of the mass-parameter amplitude $\alpha_0 \,{=}\, 0.03,\ 0.045\ 0.06$, with $\alpha_H \,{=}\, 0.085$ and $\alpha_K \,{=}\, 0$. at source redshifts $z_S \,{=}\, 0.1$ (top left), $z_S \,{=}\, 0.5$ (top right), $z_S \,{=}\, 1$ (bottom left) and $z_S \,{=}\, 3$ (bottom right).}\label{apx:Clsfracs_a0} 
\end{figure}

\begin{figure}[!h]\center
\includegraphics[scale=0.4]{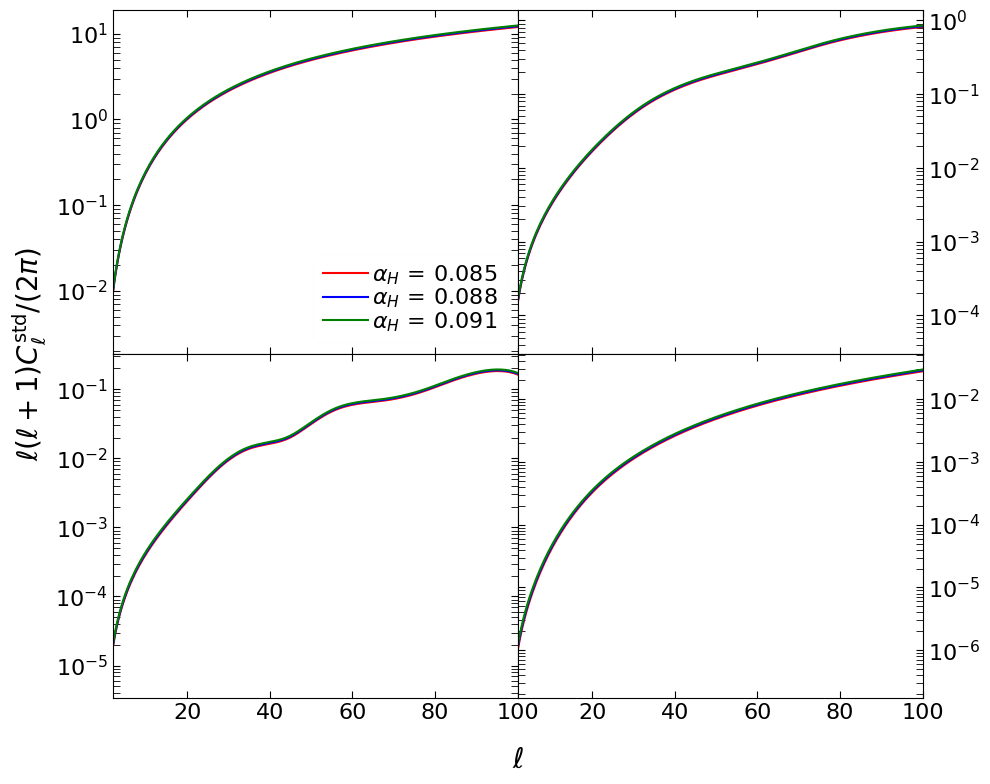}
\caption{The plots of the standard angular power spectrum, for the Horndeski parameter: $\alpha_H \,{=}\, 0.085,\ 0.088\ 0.091$, with $\alpha_0 \,{=}\, 0.06$ and $\alpha_K \,{=}\, 0$. The panel arrangements are as in Fig.~\ref{apx:Clsfracs_a0}.}\label{apx:Clsfracs_aH} 
\end{figure}

\begin{figure}\center
\includegraphics[scale=0.4]{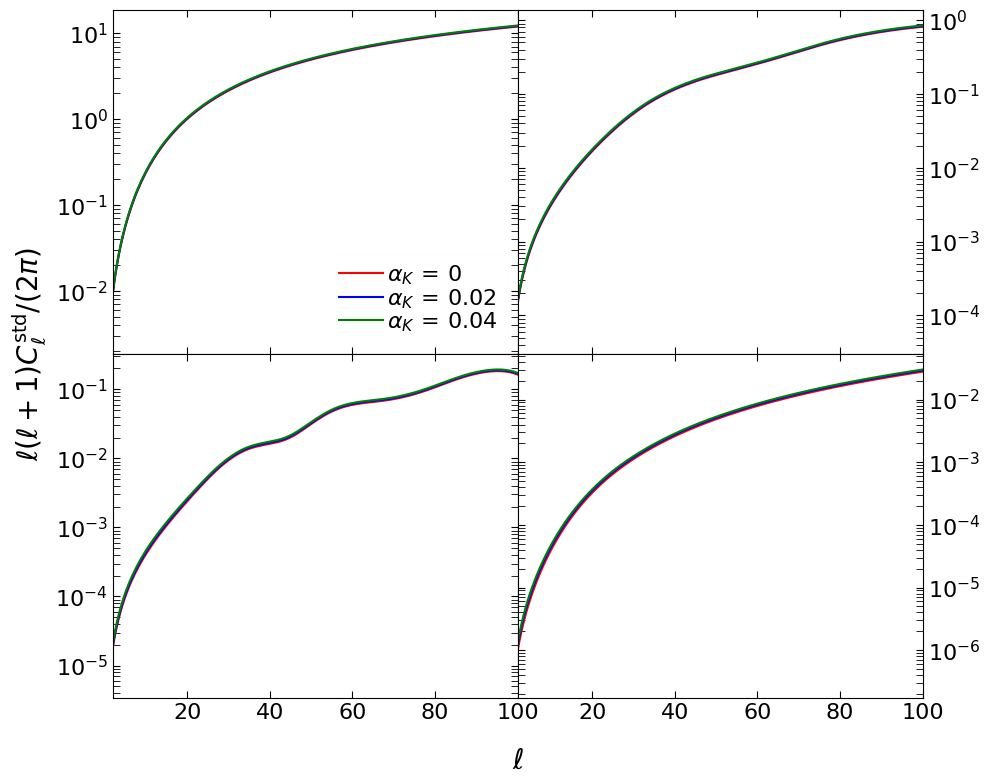}
\caption{The plots of the standard angular power spectrum, for the kineticity parameter: $\alpha_K \,{=}\, 0,\ 0.02\ 0.04$, with $\alpha_H \,{=}\, 0.085$ and $\alpha_0 \,{=}\, 0.06$. The panel arrangements are as in Fig.~\ref{apx:Clsfracs_a0}.}\label{apx:Clsfracs_aK} 
\end{figure}

In Fig.~\ref{apx:Clsfracs_a0} we see that for all the given values of $\alpha_0$, the standard angular power spectrum $C^{\rm std}_\ell$ is identical, with the various plots overlapping on each other; similarly for both $\alpha_H$ and $\alpha_K$: Figs.~\ref{apx:Clsfracs_aH} and \ref{apx:Clsfracs_aK}. This clarifies our earlier claim in the discussion of Fig.~\ref{fig:Clsfracs_aH} that on large scales the UDE parameters $\alpha_0$, $\alpha_H$ and $\alpha_K$ mainly affect the relativistic corrections. It should be noted that the change $\Delta C_\ell \,{=}\, C_\ell \,{-}\, C^{\rm std}_\ell$ will contain both the contribution of the individual relativistic terms and, also the contributions of their respective cross-terms with the standard-term components (which are the most dominant components in \eqref{Delta_n}).

\section{The Cosmological Equations}\label{sec:CEqs}
Note that all equations given in this appendix are drawn from the work by \cite{Gleyzes:2014rba}; however, we rewrite the equations with respect to conformal time. 

\subsection{The perturbations equations}
The gravitational potential are given, via the metric \eqref{metric}, by 
\beq
\Phi \;\equiv\; \delta{N} + {\cal H}\pi + \pi',\quad\quad \Psi \;\equiv\; -\zeta - {\cal H}\pi,\quad\quad \pi \;=\; a\psi,
\eeq
where $\pi = a^{-1}\pi^{(\rm phys)}$ is the comoving component, with $\psi$ being a metric scalar potential; $\delta{N}$ is the metric temporal perturbation and $\zeta$ is a metric spatial potential---and the superscript ``phys'' denotes the physical quantities as given by \cite{Gleyzes:2014rba}. The gravitational potential equations are given in subsection \ref{sec:PEqs}. The evolutions of the UDE momentum density and (energy) density perturbation, are given by (which were used in obtaining \eqref{VxEvoln} and \eqref{DxEvoln}) 
\bea
q'_x + 4{\cal H}q_x + \left(\bar{\rho}_x + \bar{p}_x\right)\Phi  + \delta{p}_x - \dfrac{2}{3} k^2 \sigma_x &\;=\;& \alpha_M {\cal H}\dsum_A{q_A},\\
\delta{\rho}'_x + 3{\cal H}\left(\delta{\rho}_x + \delta{p}_x\right) -3\left(\bar{\rho}_x + \bar{p}_x\right) \Psi' - k^2q_x  &\;=\;& \alpha_M {\cal H} \dsum_A{\delta{\rho}_A},
\eea %
where $q_A$ and $\sigma_A$ are given by \eqref{PsiDot} and \eqref{PsiPhi}, respectively. The UDE perturbations are
\bea\label{V_x}
V_x &\equiv & -\pi - \dfrac{2\alpha_B{\cal H}}{\bar{\rho}_x+\bar{p}_x} {\cal P},\\ \label{sigma_x}
\sigma_x &\equiv & \alpha_M a^{-2}M^2 {\cal H}\pi - \alpha_T {\cal R} -\alpha_H {\cal P},\\ \label{delP_x}
\delta{p}_x &\equiv & \left[\bar{p}'_x + \alpha_M {\cal H} a^{-2}M^2 \left(2{\cal H}' + {\cal H}^2\right) \right]\pi - 2\alpha_M {\cal H} {\cal Q} + \dfrac{2}{3} k^2\sigma_x \nn
&& + \left(\dfrac{\bar{\rho}_x+\bar{p}_x}{a^{-2}M^2} + 6\alpha_B {\cal H}^2\right) {\cal P} + 2\alpha_B \left(1 + \dfrac{\alpha'_B}{{\cal H} \alpha_B} + \dfrac{{\cal H}'}{{\cal H}^2} + \dfrac{{\cal P}'}{{\cal H} {\cal P}}\right) {\cal H}^2 {\cal P},\\ \label{delRho_x}
\delta{\rho}_x &\equiv & 2\left( \alpha_H{\cal R} - \alpha_B a^{-2} M^2{\cal H}\pi\right) k^2 - 3{\cal H}\left[ (\bar{\rho}_x+\bar{p}_x)\pi - 2\alpha_B {\cal Q}\right] + \left(\alpha_K - 6\alpha_B\right) {\cal H}^2 {\cal P},\quad
\eea
where
\beq\label{PQR}
{\cal P} \equiv \dfrac{M^2}{a^2} \left(\pi' + {\cal H}\pi - \Phi\right),\;\; {\cal Q} \equiv \dfrac{M^2}{a^2} \left[\Psi' + {\cal H}\Phi + ({\cal H}' - {\cal H}^2)\pi\right],\;\; {\cal R} \equiv \dfrac{M^2}{a^2} \left(\Psi + {\cal H}\pi\right),
\eeq
with the parameters related to their physical counterparts by ${\cal P} = a^{-2} {\cal P}^{(\rm phys)}$, ${\cal Q} = a^{-1} {\cal Q}^{(\rm phys)}$ and ${\cal R} = a^{-2} {\cal R}^{(\rm phys)}$.

\subsection{The metric potentials evolution equations}

The evolution equations for the metric potentials $\pi$ and $\Psi$ are given by
\bea\label{dpidt}
\pi' + \Big(1 + \dfrac{\alpha_T - \alpha_M}{\alpha_H}\Big) {\cal H}\pi &=& \Big(\dfrac{1+\alpha_H}{\alpha_H}\Big) \Phi - \Big(\dfrac{1+\alpha_T}{\alpha_H}\Big) \Psi + \dfrac{a^2\sigma_m}{\alpha_H M^2},\\ \label{dPsidt}
\Psi' + (1+\alpha_B) {\cal H}\Phi  &=& \alpha_B {\cal H}\pi' + \left(1 + \alpha_B -\dfrac{{\cal H}'}{{\cal H}^2} - \dfrac{\bar{\rho}_m + \bar{p}_m}{2a^{-2}M^2{\cal H}^2}\right) {\cal H}^2\pi - \dfrac{a^2 q_m}{2M^2},\quad
\eea
where by using \eqref{V_x}, and the expression for ${\cal P}$ given by \eqref{PQR}, in \eqref{dPsidt} we get \eqref{PsiDot}. 

The second order evolution of $\pi$ is given by
\bea\label{ddpiddt}
\pi'' + \left(1 + \gamma_1\right) {\cal H}\pi' + \gamma_3 {\cal H}^2 \pi = \Phi' - \gamma_4\Psi' - \gamma_5 {\cal H}\Phi - \gamma_6 {\cal H}\Psi - 2\gamma_7 {\cal H} \dfrac{a^2\sigma_m}{M^2} - 3\gamma_8 \dfrac{a^2\delta{p}_m}{M^2{\cal H}},\quad\quad
\eea
where have defined the parameters
\bea\label{g0}
\gamma_0 &\equiv & \alpha_K + 6\alpha_B^2,\\
\gamma_1 \gamma_0 &\equiv & (3+\alpha_M)\gamma_0 + \dfrac{\alpha'_K}{{\cal H}} + (6\alpha^2_B + 2\alpha_K - 6\alpha_B) \Big[\dfrac{{\cal H}'}{{\cal H}^2} - 1\Big] + 6\alpha_B\Big[\dfrac{\alpha'_B}{{\cal H}} - \dfrac{\bar{\rho}_m + \bar{p}_m }{2a^{-2}M^2{\cal H}^2}\Big],\quad\quad\\
\gamma_2 \gamma_0 &\equiv & - 3\alpha_B \dfrac{a^3\bar{p}'_m}{M^2{\cal H}^3} + 6\left[\dfrac{\alpha'_B}{{\cal H}} + (1+\alpha_B) \Big(\dfrac{{\cal H}'}{{\cal H}^2} - 1\Big) + \dfrac{\bar{\rho}_m + \bar{p}_m}{2a^{-2}M^2{\cal H}^2} \right] \Big(\dfrac{{\cal H}'}{{\cal H}^2} - 1\Big) \nn
&& -\dfrac{2k^2}{{\cal H}^2} \left[1 + \alpha_T + \alpha_B(1+\alpha_B) - (1+\alpha_H)(1+\alpha_M) + (1+\alpha_B-\alpha_H) \Big(\dfrac{{\cal H}'}{{\cal H}^2} -1\Big) \right.\nn
&& \hspace{2cm} + \left.  \dfrac{\alpha'_B - \alpha'_H}{{\cal H}} + \dfrac{\bar{\rho}_m + \bar{p}_m}{2a^{-2}M^2{\cal H}^2} \right],\\
\gamma_3 &\equiv & \gamma_1 + \gamma_2 + \dfrac{{\cal H}'}{{\cal H}^2},
\eea

\bea
\gamma_4 &\equiv & 6 \gamma^{-1}_0\Big[ \dfrac{\alpha'_B}{{\cal H}} + (1+\alpha_B) \Big(\dfrac{{\cal H}'}{{\cal H}^2} - 1\Big) + \dfrac{\bar{\rho}_m + \bar{p}_m}{2a^{-2}M^2{\cal H}^2} \Big] + \dfrac{2\alpha_H}{\alpha_K + 6\alpha_B^2} \dfrac{k^2}{{\cal H}^2},\\
\gamma_5 \gamma_0 &\equiv & -(3+\alpha_M)\gamma_0 - \dfrac{\alpha'_K}{{\cal H}} + 6(1-\alpha_B) \dfrac{\alpha'_B}{{\cal H}} + 2(\alpha_H - \alpha_B) \dfrac{k^2}{{\cal H}^2}+ 3(1+\alpha_B) \dfrac{\bar{\rho}_m + \bar{p}_m}{a^{-2}M^2{\cal H}^2}\nn
&& +\; \Big[6\alpha^2_B + 2\alpha_K - 12\alpha_B -6 \Big] \Big(1 - \dfrac{{\cal H}'}{{\cal H}^2} \Big),\\ 
\gamma_6 &\equiv & \dfrac{2k^2}{\gamma_0 {\cal H}^2} \Big[\alpha_M + \alpha_H(1+\alpha_M) - \alpha_T - \dfrac{\alpha'_H}{{\cal H}} \Big],\\ \label{g8} 
\gamma_7 &\equiv & \dfrac{\alpha_B}{\alpha_K + 6\alpha_B^2} \dfrac{k^2}{{\cal H}^2},~~~ \gamma_8 \equiv \dfrac{\alpha_B}{\alpha_K + 6\alpha_B^2}. 
\eea  
Note that the various $\gamma_i$ (with $i \,{=}\, 0,1,2,\cdots$) are not the same as those in e.g. \cite{Gleyzes:2014rba, Lombriser:2015cla}.

Moreover, by taking the time derivative of \eqref{dpidt}, we use \eqref{dPsidt} and \eqref{ddpiddt} to get
\beq\label{dPhidt}
\Phi' + (1+\lambda_1) {\cal H}\Phi = \lambda_2 {\cal H}\Psi + \lambda_3 {\cal H}^2\pi - \lambda_4 \dfrac{a^2 q_m}{2M^2} + \lambda_5 {\cal H} \dfrac{a^2 \sigma_m}{M^2} - 3\lambda_6 \dfrac{a^2 \delta{p}_m}{{\cal H}M^2},
\eeq
where we defined the parameters
\bea
\lambda_1 &\equiv & \alpha_T + \alpha_H(\gamma_5 -\gamma_4) + \alpha_B(1+\alpha_T -\alpha_H\gamma_4) - \dfrac{\alpha'_H}{ {\cal H}\alpha_H} - \beta_2\Big(\dfrac{1+\alpha_H}{\alpha_H}\Big),\\
\lambda_2 &\equiv & \dfrac{\alpha'_T}{{\cal H}} - (1+\alpha_T) \dfrac{\alpha'_H}{ {\cal H}\alpha_H} -\alpha_H\gamma_6 - \beta_2\Big(\dfrac{1+\alpha_T}{\alpha_H}\Big),\\
\lambda_3 &\equiv & \beta_3 + \beta_2\Big(\dfrac{\alpha_M - \alpha_T}{\alpha_H} -1\Big),\quad \lambda_4 \equiv 1+\alpha_T - \alpha_H \gamma_4,\\ 
\lambda_5 &\equiv & \alpha_M +\dfrac{\alpha'_H}{ {\cal H}\alpha_H} - \dfrac{\sigma'_m}{ {\cal H}\sigma_m} -2\Big(1 + \alpha_H\gamma_7\Big) + \dfrac{\beta_2}{\alpha_H}, \quad \lambda_6 \equiv \alpha_H\gamma_8,
\eea
and we have
\bea
\beta_2 &\equiv & \alpha_T + \alpha_B \lambda_4 - \alpha_M - \alpha_H\gamma_1,\\
\beta_3 &\equiv & (\alpha_M -\alpha_T)\dfrac{\alpha'_H}{\alpha_H {\cal H}} + (\alpha_H -\alpha_M +\alpha_H\gamma_4 - 1)\dfrac{{\cal H}'}{{\cal H}^2} + \dfrac{\alpha'_T -\alpha'_M}{{\cal H}} \nn
&& -\alpha_H\gamma_3 + \lambda_4\Big(1+\alpha_B - \dfrac{\bar{\rho}_m + \bar{p}_m}{2a^{-2}M^2{\cal H}^2}\Big).
\eea

\section{Adiabatic Initial Conditions}\label{sec:AICs}

We use the Einstein de Sitter initial condition $\Psi'(z_d)=0$ at the decoupling epoch $z=z_d$, given that $\Omega_x(z_d)\ll 1$. Moreover, adiabatic initial conditions are usually imposed by the vanishing of the relative entropy perturbation $S_{xm}$ (see e.g. \cite{Duniya:2015ths, Duniya:2013eta, Duniya:2015nva, Duniya:2015dpa}), given by
\beq \label{adic}
S_{mx}(z_d) \;=\; 0, \quad\quad S_{mx} \;\equiv\; 3{\cal H} \left( {\delta\rho_m \over \bar{\rho}'_m} - {\delta\rho_x \over \bar{\rho}'_x}\right).
\eeq
Then by choosing the velocities to be equal, 
\beq\label{vels_d}
V_x(z_d) \;=\; V_m(z_d),
\eeq
and noting $\Delta_A$ in \eqref{Pois}, we obtain
\beq
\dfrac{\Delta_x(z_d)}{1 +w_{x,\rm eff}(z_d)} \;=\; \dfrac{\Delta_m(z_d)}{1 +w_m(z_d)} .
\eeq
Together with \eqref{Pois} and \eqref{PsiDot}, we get the initial fluctuations at $z_d$, given by
\bea \label{V_mi:V_xi}
V_{md}(k) &\;=\;& \dfrac{-2}{3{\cal H} \left(1 + \Omega_m w_m + \Omega_x w_x\right)} \Phi_d(k), \\ \label{Delta_mi}
\Delta_{md}(k) &\;=\;& \dfrac{1+w_m}{1 + \Omega_m w_m + \Omega_x w_{x,\rm eff}} \Big[\alpha_M {\cal H}V_{xd}(k) - \dfrac{2k^2}{3{\cal H}^2}\Psi_d(k)\Big], 
\eea
where $\Phi_d$ is given by \eqref{Phi_d}, and given \eqref{dpidt} and $\pi'(z_d)=0=\sigma_m(z_d)$, we have
\beq\label{Phi}
\Psi_d(k) = \dfrac{1+\alpha_H}{1+\alpha_T} \Phi_d(k) + \dfrac{\alpha_M -\alpha_T -\alpha_H}{1+\alpha_T} {\cal H}\pi_d(k),
\eeq
and from \eqref{V_x}, with ${\cal P}$ being given by \eqref{PQR}, we have
\beq\label{pi_d}
\pi_d(k) = -\dfrac{1}{\lambda_7} V_{xd}(k) + \dfrac{2\alpha_B}{3\lambda_7 (1+w_x) {\cal H}\Omega_x} \Phi_d(k) ,
\eeq
with the parameter $\lambda_7 \equiv 1 + 2\alpha_B / [3(1+w_x)\Omega_x]$. (Note that throughout this appendix, all the background parameters are also evaluated at $z=z_d$.)


\end{document}